\documentclass[aps,twocolumn,prb,showpacs]{revtex4}

\usepackage{amsmath,amsfonts,amssymb,color,graphicx}
\usepackage[colorlinks]{hyperref}
\definecolor{ginored}{rgb}{0,0,0}
\definecolor{blue}{rgb}{0,0,0}
\definecolor{green}{rgb}{0,0,0}
\definecolor{darkred}{rgb}{0,0,0}
\begin{document}

%\title{Structure and melting behavior of a classical bilayer system of dipoles}
\title{Structure and melting behavior of classical bilayer crystals of dipoles}

\author{Xin Lu$^*$ and Chang-Qin Wu}

%\author{Xin Lu$^{1}$, Chang-Qin Wu$^{1}$, Andrea Micheli$^{2,3}$ and Guido Pupillo$^{2,3}$}
%\affiliation{$^{1}$Department of Physics, Fudan University, Shanghai
%200433, China\\$^{2}$Institute for Theoretical Physics, University of Innsbruck,
%A--6020 Innsbruck, Austria \\
%$^{3}$Institute for Quantum Optics and Quantum Information of the
%Austrian Academy of Sciences, A-6020 Innsbruck, Austria}
\affiliation{Surface Physics Laboratory (National Key Laboratory)
and Department of Physics, Fudan University, Shanghai 200433,
People¡¯s Republic of China}

\author{Andrea Micheli and Guido Pupillo}

\affiliation{Institute for Theoretical Physics, University of
Innsbruck,
A--6020 Innsbruck, Austria \\
and Institute for Quantum Optics and Quantum Information of the
Austrian Academy of Sciences, A-6020 Innsbruck, Austria}

\begin{abstract}
We study the structure and melting of a classical bilayer system of
dipoles, in a setup where the dipoles are oriented perpendicular to
the {planes of the layers} and the density of dipoles is the same in
each layer. Due to the anisotropic character of the dipole-dipole
interactions, we find that the ground-state configuration is given
by two hexagonal crystals positioned on top of each other,
independent of the interlayer spacing and dipolar density. {For
large interlayer distances these crystals are independent, while in
the opposite limit of small interlayer distances the system behaves
as a two-dimensional crystal of paired dipoles. Within the harmonic
approximation for the phonon excitations, the melting temperature of
these crystalline configurations displays} a non-monotonic
dependence on the interlayer distance, which is associated with a
{\em re-entrant} melting behavior in the form of {\it
solid-liquid-solid-liquid} transitions at fixed temperature. %Our
%analysis makes use of the harmonic approximation for the excitations
%of the crystal, where rapidly convergent forms for the static and
%dynamic properties of the system are obtained via an Ewald summation
%method.
\end{abstract}


\maketitle

\section{Introduction}

{The realization of a degenerate dipolar gas of $^{52}$Cr atoms
\cite{pfau1,pfau2} and the experimental progress in the realization
of cold molecular ensembles \cite{Exp} have spurred interest in the
properties of particles with {large dipole moments
\cite{Book,col1,col2,col4,col5,col6,all7,all10,Micheli07} in atomic
and molecular setups \cite{excitons}.} In particular, the strong
anisotropic dipole-dipole interactions induced in ground-state polar
molecules by external electric fields hold promises for applications
ranging from the realization of {novel strongly-correlated phases
\cite{all1,all2,all3,all4,all5,Gorshkov08,mm1,mm2,mm3,col7,Buechler07,Astrakharchik07,Mora07,Ark1,all6,all13,Opt1,Opt2,Opt3,Opt4,Opt5}
to  quantum simulations \cite{micheli06,brennenNJP07,Pupillo08} and
quantum} computing \cite{all8,all9,all6,all11,all12,all14}.}

\begin{figure}[tbp!]
\begin{center}
\includegraphics[width=.9\columnwidth]{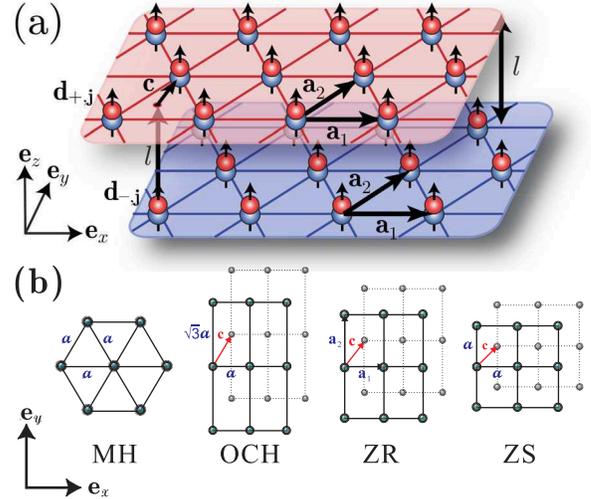}
\caption{\label{fig:fig01:lattice}(Color online) (a) Sketch of the
system setup depicting the dipolar bi-layer system: The dipoles are
oriented along the $z$ direction, ${\bf d}=d {\bf e}_z$, and
confined to two $(x,y)$ planes separated by an interlayer distance
$l$ along the $z$ direction. The intralayer dipole-dipole repulsion
gives rise to a crystalline structure in each layer with basis
vectors ${\bf a}_i$, here illustrated by two hexagonal structures
with an in-plane displacement ${\bf c}$. (b) Various possible
equilibrium geometries for the bilayer crystal, shown as a
projection onto the $(x,y)$ plane. From left to right: a matching
hexagonal (MH) configuration, a one-component hexagonal (OCH)
configuration, a zigzag rectangular configuration with aspect ratio
$a_2/a_1$ (ZR), and a zigzag square (ZS) configuration. The position
of the dipoles and their nearest-neighbor links in the upper (lower)
layer are shown by solid (dashed) circles and lines, respectively. A
detailed summary of the
lattice parameters is given in Table \ref{tab:tab1}.}% The MH
%configuration is the ground-state for all $l$ and dipole densities,
%while the OCH, ZR and ZS configurations are metastable excited-state
%configurations, see text.}
\end {center}
\end{figure}

{In Refs.~\cite{Buechler07,Micheli07}} it is shown that a
collisionally stable two-dimensional (2D) setup where particles
interact via {\it purely repulsive} effective dipole-dipole
interactions can be realized by polarizing the molecules using an
external electric field, and thus inducing strong dipole-dipole
interactions, and by confining the motion {of the particles} to a 2D
geometry, {e.g. by trapping them into} {a single well of} a deep
optical lattice directed parallel to the electric field. The
low-energy phase of an ensemble of interacting bosonic polar
molecules is then a superfluid or a self-assembled crystal for
comparatively weak and strong interactions, respectively, where the
strength of the interactions can be modified by varying the
intensity of the polarizing {electric} field. The crystal is a
two-dimensional hexagonal lattice structure with quantum dynamics
given by longitudinal and transverse acoustic phonons.
Contrary to familiar Wigner crystals induced by strong Coulomb
interactions \cite{ions}, dipolar crystals emerge at large
densities, where dipole-dipole interactions dominate over the
kinetic energy of the particles. This scenario for the realization
of 2D crystals can be implemented using closed-shell polar molecules
as, e.g., SrO or RbCs.

While the scenario above is realized by populating a single well of
the confining optical lattice, in general it will be possible {to
populate more than one well of the lattice.} It is thus natural to
consider the phases of bi- and multilayer configurations of
classical and quantum dipoles. As a first step, in this work we
focus on the crystalline structures of a bilayer system of classical
dipoles. The analogous quantum problem will be the subject of a
separate study.
\\

In this paper we consider a system of $2N$ dipoles confined into two
parallel two-dimensional planes along the $(x,y)$-directions,
separated by an interlayer spacing $l$ along $z$. Each plane has the
same number of dipoles, $N$, with their dipole-moment ${\bf d}=d{\bf
e}_z$ aligned perpendicular to the plane, see
Fig.~\ref{fig:fig01:lattice}(a). %Such a setup could be realized, e.g., by
%polar molecules confined in an one-dimensional optical lattice and
%aligned by an external electric field.
The interactions between two dipoles separated by ${\bf r}$ are the
dipole-dipole interactions $V({\bf r})=d^2[1-3z^2/r^2]/r^3$, where
$r=|{\bf r}|>0$
and $z={\bf r}\cdot{\bf e}_z$. %We note that the dipole-dipole
%interactions are long-range, i.e. decaying like $1/r^3$ and
%anisotropic in space.
For our setup this results in the intra-layer interactions being
always repulsive, which {gives rise to a hexagonal crystalline
structure} in each layer. In contrast, the inter-layer interactions
are attractive (repulsive) for {dipoles} separated by a distance
$r<\sqrt{3}l$ ($r>\sqrt{3}l$), see Fig.~\ref{fig:fig01:lattice}(a).
As detailed in Secs. II-IV, this anisotropy in the inter-layer
interactions {determines that the groundstate configuration of the
system is a bilayer crystal comprised of two hexagonal crystals
stacked on top of each other. In addition, it} gives rise to a
non-monotonic behavior of the dynamic properties {of this bilayer
crystal (such as the phonon sound velocities and the melting
temperature)} when increasing the interlayer distance $l$, {while
keeping the intra-layer densities $n$ fixed.} In particular, within
the harmonic approximation for the phonon excitations we find that
for certain fixed temperatures the system displays a {\em reentrant}
melting behavior in the form of {\it solid-liquid-solid-liquid}
transitions as a function of the dimensionless inter-layer distance
{$\xi = l \sqrt{n}$.} The picture emerging from these studies is one
where for large interlayer separations {$\xi\gg 1$} the two
hexagonal crystals of the bilayer structure behave as independent
crystals, while for vanishing inter-layer separations {$\xi\ll 1$}
they behave as a single 2D crystal of particles with double mass and
double dipole-moment.
\\

The paper is organized as follows: In Section~\ref{sec:sec2}  we
introduce our model and derive expressions for the potential energy
in the form of a rapidly convergent sum obtained via the Ewald
summation method. By comparing the energies of a number of possible
crystal geometries, we determine the structure and energy of the
ground-state and of several excited-state configurations. In
Section~\ref{sec:sec3} we analyze the dynamic properties
corresponding to these crystalline structures within the harmonic
approximation for {the phonon excitations}. {We discuss the phonon
excitation branches of the ground-state configuration} and we find
that the dependence of the sound-velocities on the inter-layer
separation $l$ is non-monotonic, due to the anisotropic character of
the dipole-dipole interaction. For the excited-states we assess the
stability of the configurations studied in Sec.~\ref{sec:sec2} under
phonon fluctuations and find that several {of the excited states are
meta-stable. In Section~\ref{sec:sec4} we study the thermal melting
of the bilayer-crystal.}
%in both the ground and meta-stable
%excited states.
We derive the classical melting temperature $T_{\rm
m}$ via a modified two-dimensional Lindemann criterion, and find
that the temperature $T_{\rm m}$ behaves non-monotonically with
increasing inter-layer distance.

\section{Ground and excited states}\label{sec:sec2}

We consider a system of {\em classical} dipoles confined into two
planes along  the $x-y$ direction, separated by a distance $l$ along
$z$. We focus on the situation where {the dipole-moments} ${\bf
d}_i$ of the particles are aligned perpendicular to the planes, i.e.
along $z$, and where one has the same density $n$ of dipoles in each
layer, see Fig.~\ref{fig:fig01:lattice}(a). The interactions between
two dipoles is given by their dipole-dipole interactions $V({\bf
r})=d^2[1-3z^2/r^2]/r^3$, where $r=|{\bf r}|>0$ is their distance
and $z={\bf r}\cdot{\bf e}_z$ their interlayer distance. We note
that the dipole-dipole interactions are long-range, i.e. decaying
like $1/r^3$ and anisotropic in space. At zero-temperature the
system is in a crystalline configuration, where the particles in
each layer form a 2D crystal, and the relative position of the
particles is correlated by the long-range dipole-dipole interaction.
We denote the 2D position of the dipoles within the upper (lower)
layer by ${\bf R}_{+,j}$ (${\bf R}_{-,j}$), which we parameterize by
\begin{align}
{\bf R}_{\pm,j}=j_1{\bf a}_1+j_2{\bf a}_2\pm{\bf c}/2,
\end{align}
where the {integers $j\equiv(j_1,j_2)$ label} the $j^{\rm th}$
particle in each layer, ${\bf a}_1$ and ${\bf a}_2$ are the basis
vectors of the periodic structure {with density $n$} and ${\bf c}$
is a two-dimensional vector accounting for a relative in-plane
displacement of the two {structures}, see
Fig.~\ref{fig:fig01:lattice}(a).

The interactions are given by
\begin{align}\label{eq:V}
V =~& \frac{1}{2} \sum_{\sigma=\pm}\sum_{j\neq{j'}}
\frac{d^2}{|{\bf R}_{\sigma,j}-{\bf R}_{\sigma,j'}|^3}+\nonumber\\
&\sum_{j,j'}\frac{d^2 \left(|{\bf R}_{+,j}-{\bf
R}_{-,j}|^2-2l^2\right)}{\left(|{\bf R}_{+,j}-{\bf
R}_{-,j}|^2+l^2\right)^{5/2}},
\end{align}
where the first (second) term describes the intra-layer
(inter-layer) interactions. Since the intra-layer interactions do
not depend on the inter-layer separation $l$, it is convenient to
split the energy per dipole, $E$, into its intra- and inter-layer
part as $E=V/2N=E_0 + E_I$. Making use of the translational
invariance of the infinite system the two contributions read
\begin{subequations}\label{eq:e0ei1}
\begin{align}
E_0=&\frac{1}{2}\sum_{j\neq0}\frac{d^2}{|{\bf R}_j|^3}, \\
E_I=&\frac{1}{2}\sum_j\frac{d^2 \left(|{\bf R}_j+{\bf
c}|^2-2l^2\right)}{\left(|{\bf R}_j+{\bf c}|^2+l^2\right)^{5/2}},
\end{align}
\end{subequations}
%
%
%\eqref{eq:e0ei1},\eqref{eq:e0a},\eqref{eq:e0b}
%
%
where {${\bf R}_j\equiv{\bf R}_{\sigma,j}-{\bf R}_{\sigma,0}=j_1{\bf
a}_1+j_2{\bf a}_2$ denote the relative (2D) positions} of the
dipoles in each layer. Given the slow convergence of the sum in real
space involved  in Eq.~\eqref{eq:e0ei1}, we use the Ewald summation
method to obtain an expression for $E$ involving rapidly convergent
sums. The explicit derivation is given in Appendix~\ref{sec:appA}.
We here {provide only the derived} expressions for $E_0$ and $E_I$,
which read
\begin{subequations}\label{eq:4}
\begin{align}
\frac{E_0}{d^2} =&  \pi n \sum_{j} \left[ \frac{4\alpha}{\sqrt{\pi}} e^{-|{\bf G}_j|^2/4\alpha^2}%
- 2 |{\bf G}_j| {\rm erfc} \left(\frac{|{\bf G}_j|}{2 \alpha} \right) \right]\nonumber\\%
&+\sum_{j\neq0}  \left[ \frac{{\rm erfc}(\alpha |{\bf R}_j|)}{|{\bf R}_j|^3} %
+ \frac{2 \alpha}{\sqrt{\pi}}\frac{e^{-\alpha^2 |{\bf R}_j|^2}}{|{\bf R}_j|^2} \right]%
-\frac{4 \alpha^3}{3 \sqrt{\pi}}, \label{eq:e0rap}\\%
\frac{E_I}{d^2} =& \pi n \sum_{j} e^{i {\bf G}_j \cdot {\bf c}}\left[%
\frac{4 \alpha}{\sqrt\pi} e^{-|{\bf G}_j|^2/4\alpha^2 - \alpha^2 l^2}\right.\nonumber\\%
&\left. - |{\bf G}_j|\sum_\pm e^{\pm|{\bf G}_j|l}{\rm erfc}\left(\frac{|{\bf G}_j|}{2 \alpha}\pm\alpha l \right)\right] +\nonumber\\%
 & \sum_{j} \left[ \frac{{\rm erfc}\left( \alpha |\tilde{\bf R}_j| \right)}{|\tilde{\bf R}_j|^3}\left(1-\frac{3l^2}{|\tilde{\bf R}_j|^2} \right) \right.\nonumber\\%
 &\left.+  \frac{2 \alpha}{\sqrt{\pi}}  \frac{e^{-\alpha^2 |\tilde{\bf R}_j|^2}}{|\tilde{\bf R}_j|^2}\left(1-\frac{3l^2}{|\tilde{\bf R}_j|^2}-2\alpha^2|\tilde{\bf R}_j|^2\right)\right],\label{eq:eirap}
\end{align}
\end{subequations}
with $|\tilde{\bf R}_j|\equiv(|{\bf R}_j+{\bf c}|^2+l^2)^{1/2}$. {In
Eq.~\eqref{eq:4} $n$ denotes the intra-layer density and ${\bf G}_j$
are the 2D reciprocal vectors of a single layer, which are}
parametrized by ${\bf G}_j=j_1 {\bf b}_1+j_2 {\bf b}_2$ in terms of
the primitive translation vectors of the reciprocal lattice ${\bf
b}_1$ and ${\bf b}_2$. The quantity $\alpha>0$ is an (arbitrary)
inverse length, for which a convenient choice is the inverse of the
mean particle separation, i.e., $\alpha = 1/r_0 = \sqrt{\pi n}$.

\begin{table}[tbp!]
\caption{\label{tab:tab1}Lattice parameters of four considered
configurations: the matching hexagonal (MH), the one-component hexagonal (OCH), the zigzag rectangular
(ZR) and a zigzag square (ZS). From left to right: ${\bf a}_1$ and ${\bf a}_2$ are the primitive vectors,  ${\bf c}$ is the inter-lattice displacement
vector, ${\bf b}_1$ and ${\bf b}_2$ are the primitive translation vectors of
the reciprocal lattice and $n$ is the density of each layer. We introduced the vector notation $(x, y) \equiv x {\bf e}_x + y {\bf e}_y
$ for the two-dimensional components; the lattice constant is $a\equiv|{\bf a}_1|$ and $a_2/a_1$ denotes the aspect ration for ZR configuration.}
\begin{center}
\begin{tabular}{ccccccccccccc}
lattice && ${\bf a}_1/a$ &&  ${\bf a}_2/a$ &&  $2{\bf c}$ &&  ${\bf b}_1 a/2\pi$ && ${\bf b}_2a/2\pi$ && $n a^2$ \\
\hline
 MH    && $(1,0)$ && $(\frac{1}{2},\frac{\sqrt3}{2})$ && $ 0 $                       && $(1,\frac{-1}{\sqrt{3}})$ && $(0,\frac{2}{\sqrt{3}})$ && $\frac{2}{\sqrt{3}}$  \\
 OCH   && $(1,0)$ && $(0,\sqrt{3})$     && $(1,\sqrt3) $ && $(1,0)$           && $(0,\frac{1}{\sqrt{3}})$ && $\frac{1}{\sqrt{3}}$  \\
 ZS    && $(1,0)$ && $(0,1)$            && $(1,1)$ && $(1,0)$           && $(0,1)$          && $1$  \\
 ZR    && $(1,0)$ && $(0,\tfrac{a_2}{a_1})$      && $(1,\frac{a_2}{a_1})$ && $(1,0)$           && $(0,\frac{a_1}{a_2})$    && $\frac{a_1}{a_2}$  \\
\end{tabular}
\end{center}
\end{table}

In order to determine the ground-state configuration, in the
following we calculate the energy of a number of possible crystal
configurations.

\begin{figure}[tbp!]
\includegraphics[width=\columnwidth]{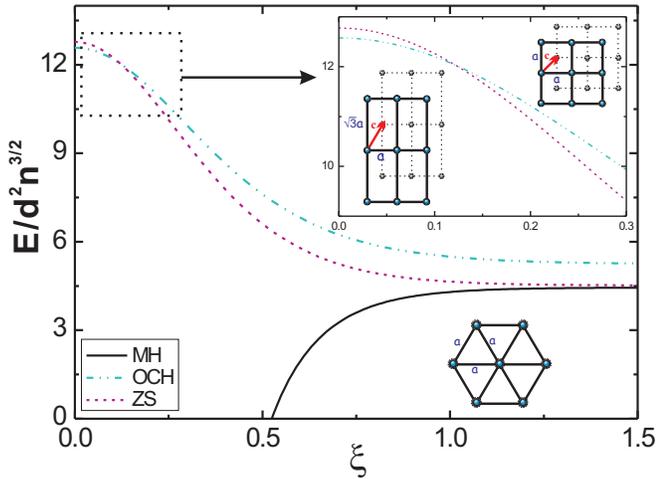}
\caption{(Color online) The energy per dipole $E$ for a fixed
density $n$ as a function of $\xi$ for three different lattice
configurations: MH (solid line), OCH (dash-dotted line) and ZS
(dotted line). The inset is a blow-up of the excited state energies
in the region of small interlattice separation, $0\leq\xi\leq0.3$,
showing a crossing of the OCH- and ZS-energies around
$\xi\approx0.12$.} \label{fig:fig02:entot}
\end{figure}

{Motivated by the fact} that the ground-state for a single layer is
given by a hexagonal lattice structure, we first consider a
hexagonal structure in each layer with the two structures displaced
by ${\bf c}$, see Fig.~\ref{fig:fig01:lattice}(a). We find that for
an arbitrary inter-layer separation $l$, the minimal energy is
attained for ${\bf c}=0$ (modulo the lattice constant). This is
{also what one would intuitively expect}, since the closest dipoles
in different layers tend to attract each other, which leads to a
locking of the relative positions of the two layers. The attained
``matching'' hexagonal (MH) structure is illustrated in
Fig.~\ref{fig:fig01:lattice}(b) and its basis and reciprocal vectors
are listed in Table \ref{tab:tab1}, together with {three} other
(meta-stable) configurations detailed below. In
Figure~\ref{fig:fig02:entot} we plot the energy per particle
{$E^{({\rm MH})}$  attained for the MH configuration} at a fixed
density $n$ as a function of the dimensionless inter-layer
separation {$\xi{\equiv}l/(\sqrt{\pi}r_0)$,} given by the
inter-layer distance in units of the mean particle separation (up to
a constant $1/\sqrt{\pi}$), {as a solid line}. We notice that for
large inter-layer separations, $\xi \gg 1$, the ground-state energy
approaches the value corresponding to a single hexagonal (SH) layer
in its ground-state, i.e. $E^{({\rm MH})}(\xi \rightarrow \infty)=
E^{({\rm SH})}\approx 4.443 d^2 n^{3/2}$. In the opposite limit,
$\xi \ll 1$, the energy is dominated by the large attraction between
the two layers {$E^{(\rm MH)}(\xi \ll 1)\approx (-1/\xi^3 + 2\times
4.443) d^2 n^{3/2}$,} {which diverges as $\propto -1/l^3$ for
$\l\rightarrow 0$, due to the strong attraction of the closest
dipoles in different layers}. We remark that the latter corresponds
to an unphysical regime for typical molecular
{systems~\cite{Micheli07}}. In fact, at these short distances the
required {depth of the optical trapping potential for realizing a 2D
layer becomes unreasonably large. Moreover} one expects short-range
interactions in the form of, e.g., Van-der-Waals interactions as
well as the core repulsion,
to dominate the {interactions at these small distances~\cite{Groh}}. These latter effects have been neglected in our {model, c.f.} Eq.~\eqref{eq:V}. {However, we notice that here the physically relevant quantity is $\xi$, and the limit $\xi \rightarrow 0$ can be approached by decreasing the density of dipoles $n$ while keeping $l$ fixed.}\\%We notice that in

In order to confirm that the MH configuration is the ground-state,
we compare its energy to those of a number of other (intuitively
motivated) configurations \cite{goldoni}.

{As a first alternative configuration} we consider a ``one
component'' hexagonal (OCH) structure, which is obtained by removing
every second dipole in each layer in a staggered way and rescaling
{the} relative density. The OCH configuration is illustrated in
Fig.~\ref{fig:fig01:lattice}(b), and its basis and reciprocal
vectors are summarized in Table \ref{tab:tab1}. {The attained energy
$E^{(\rm{OCH})}$ is plotted in Fig.~\ref{fig:fig02:entot} as a
function of $\xi$ as a dashed-dotted line, and we notice that it
exceeds that of the MH configuration. The label OCH has been chosen
for this configuration, as it {resembles} a hexagonal lattice for a
single component, when looked from above. Accordingly}, in the limit
of vanishing inter-lattice separation the energy of the OCH
configuration tends to that of a single hexagonal layer {(with a
double density), i.e. $E^{({\rm OCH})}(\xi=0)\approx
4.443~d^2(2n)^{3/2}\approx12.576~d^2 n^{3/2}$}.

{As a second alternative configuration we consider} a ``zig-zag''
square (ZS) structure, where dipoles in each layer form a square
lattice, and the two layers are shifted with respect to each other
as illustrated in Fig.~\ref{fig:fig01:lattice}. {The basis and
reciprocal {vectors} for the ZS structure are given in Table
\ref{tab:tab1}}. The energy for the ZS-structure $E^{({\rm ZS})}$ is
shown in Fig.~\ref{fig:fig02:entot} as a dashed line. The figure
shows that for large $\xi$ the energy of the ZS configuration
exceeds the MH energy but it is smaller than the OCH energy, while
at small $\xi$ the energies of the ZS and OCH configurations become
comparable. The inset of Fig.~\ref{fig:fig02:entot} is a blow-up of
the excited state energies in the region $0\leq\xi\leq0.3$. From the
inset, we observe that the {two energies, $E^{({\rm OCH})}$ and
$E^{({\rm ZS})}$}, actually cross at $\xi=\xi_0\approx 0.12$ and for
$\xi<\xi_0$ the OCH structure is energetically favored over the ZS
one.

\begin{figure}[tbp!]
\begin{center}
\includegraphics[width=\columnwidth]{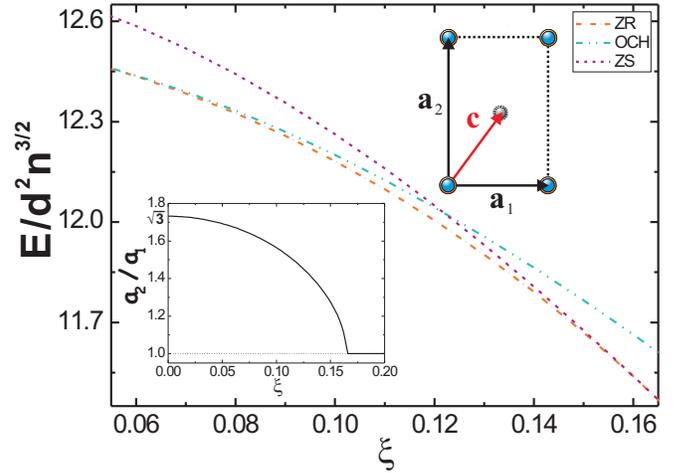}
\caption{\label{fig:fig03:enzr}(Color online) The energy {per
particle} $E$ as a function of $\xi$ {for the OCH (dash-dotted
line), the optimal ZR (dashed line) and ZS configuration (dotted
line)}, showing the transitions: $\textrm{OCH} \rightarrow
\textrm{ZR} \rightarrow \textrm{ZS}$ {with increasing $\xi$}. Inset:
The aspect ratio $a_2/ a_1$ for the ZR-configuration as a function
of $\xi$.}
%The vectors ${\bf a}_1$, ${\bf a}_2$ and ${\bf c}$ are given in
%Table \ref{tab:tab1}.
\end{center}
\end{figure}

The excited state configurations OCH and ZS can be interpolated
smoothly by "stretching" the lattice, which in general leads to a
``zig-zag'' rectangular (ZR) structure with a variable aspect ratio
$a_2/a_1$ in each layer. The ZR structure is illustrated in
Fig.~\ref{fig:fig01:lattice}(b) and its basis and reciprocal vectors
are listed in Table~\ref{tab:tab1}. In order to find the optimal ZR
configuration, we minimize the energy as a function of the free
parameter $a_2/a_1$. {The obtained aspect ratio $a_2/a_1$ for the
optimal ZR configuration in shown in the inset of
Fig.~\ref{fig:fig03:enzr}, and we notice that the resulting
structure {coincides} with the OCH and ZS configurations at $\xi=0$
and $\xi\approx0.17$, respectively. The energy obtained for the
optimal ZR configuration, $E^{({\rm ZR})}$, is plotted as a dashed
line in Fig.~\ref{fig:fig03:enzr} in {the} region
$0.05\lesssim\xi\lesssim0.17$ , along with the ones obtained for the
OCH (dash-dotted line) and ZS structures (dotted line). {The figure
shows that $E^{({\rm ZR})}$ equals $E^{({\rm OCH})}$ and $E^{({\rm
ZS})}$ at $\xi=0$ and $\xi\approx0.17$, respectively, while it is
lower in the parameter region in between.}}

The results above are consistent with the MH configuration being the
ground-state of the system for all $\xi$, due to the attraction
between particles in the two layers {(separated by $r<\sqrt{3}l$)}.
This in contrast to the situation occurring for Coulomb bilayer
systems, where the repulsion between particles in the different
layers leads to a change of the ground-state configuration depending
on the ratio of the inter-layer separation to the mean intra-layer
spacing \cite{goldoni}. However, we notice that analogous (smooth)
transitions between different configurations occur here between
excited-state structures, which will be shown below to be
metastable. These structures may in principle be prepared using
properly-designed in-plane optical lattice potentials, as argued
below.

%
%

%%%%%%%%%%%%%%%%%%%%%%%%%%%%%%%%%%%%%%%%%%%%%%%%%%%%%%%%

\section{Dynamical properties}\label{sec:sec3}

In this section, we study the phonon spectra for those equilibrium
lattice configurations, which we found in the previous section.
Thereby we make use of the harmonic approximation of the lattice
excitations, and determine the stability (meta-stability) of the
obtained configurations for the ground (excited states) under small
fluctuations. %Moreover, the obtained phonon spectra provide us also information about \todo{ldots}.

In order to obtain the excitation spectra for the various lattice configurations, we consider the dynamical matrix  ${\bf M}_{\rm D}({\bf q})$, whose eigenvalues are the square of the phonon frequencies~\cite{maradudin}.
We notice that the bilayer configurations above correspond to a
(single 2D) Bravais lattice with a unit cell given by two molecules,
one in the upper and one in the lower layer. Thus ${\bf M}_{\rm
D}({\bf q})$ is here a $4 \times 4$ matrix
\begin{eqnarray}
{\bf M}_{\rm D}({\bf q}) = \left(
\begin{array}{llll}
M_{++}^{xx} & M_{++}^{xy} & M_{+-}^{xx} & M_{+-}^{xy}  \\
M_{++}^{xy} & M_{++}^{yy} & M_{+-}^{xy} & M_{+-}^{yy}  \\
M_{+-}^{xx} & M_{+-}^{xy} & M_{--}^{xx} & M_{--}^{xy}  \\
M_{+-}^{xy} & M_{+-}^{yy} & M_{--}^{xy} & M_{--}^{yy} \\
\end{array}
\right),\label{eq:eqDynMatr}
\end{eqnarray}
where the superscript $\tau=x$ ($\tau=y$) refer to a displacement %dislocation
along $x$ ($y$) and the subscript $\sigma=+$ ($\sigma=-$) refer to the components in the upper (lower) layer. The matrix elements in Eq.~\eqref{eq:eqDynMatr} read
\begin{subequations}
\begin{align}
{M_{\sigma\sigma}^{\tau\nu}  \equiv}& ~\frac{1}{m} \left[
D_{++}^{\tau\nu}(0)
  - D_{++}^{\tau\nu}({\bf q}) + D_{+-}^{\tau\nu}(0) \right], \\
M_{+-}^{\tau\nu} \equiv& ~ \frac{1}{m} [ -D_{+-}^{\tau\nu}({\bf
q}) ],\label{eq:MM}
\end{align}
\end{subequations}
where $m$ is the mass of the dipoles. % and we made use of the fact
%that the two layer have the same structure.
The quantities $D_{+\sigma}^{\tau\nu}({\bf q})$ are defined as
{\begin{eqnarray} D_{+\sigma}^{\tau\nu} ({\bf q})  &=& \sum_j e^{ -i
{\bf q} \cdot ({\bf R}_{0,+}-{\bf
R}_{j,\sigma})}\partial_\tau\partial_\nu V_{0+,j\sigma}({\bf
r}=0),\label{eq:7}
% D_{+\sigma}^{\tau\nu} ({\bf q})  &=&  \Big[\frac{\partial^2}{\partial_{\tau}
%\partial_{\nu}} \sum_j V_{0+,j\sigma}(\bf r) \Big] {\rm e}^{ -i {\bf q} \cdot {\bf
%R}_{0+,j\sigma} } \label{eq:7}\end{eqnarray}
\end{eqnarray}}
with $V_{0+,j\sigma}(\bf r)$ the two-body interaction potential
between the dipole at position $0$ in layer $+$ and {the} dipole $j$
in layer $\sigma$. {At ${\bf q}=0$ the quantities
$D_{+\sigma}^{\tau\nu}({\bf q})$ for $\sigma=+$ ($\sigma=-$)
correspond to the intra-layer (inter-layer) force constants.}

Using the Ewald summation method [see
Eq.~\eqref{eq:psi0rap} and Eq.~\eqref{eq:psiirap} in
Appendix~\ref{sec:appA}], we can
rewrite the sum in Eq.~\eqref{eq:7} into the following rapidly convergent forms
\begin{subequations}
\begin{align}
D_{++}^{\tau\nu}({\bf q}) =&  -\sum_{j} ({\bf q} + {\bf
G}_j)_{\tau}({\bf q} + {\bf G}_j)_{\nu} \Upsilon\left(
\tfrac{|{\bf G}_j + {\bf q}|}{2 \alpha}, 0 \right) \nonumber \\
& + \frac{8 \alpha^5}{5
 \sqrt{\pi}} \delta_{\tau\nu} + \sum_{{j} \ne 0} \lim_{{\bf r} \rightarrow 0}
\partial_{\tau}\partial_{\nu} \Omega_1 \big( |{\bf R}_j + {\bf r}| \big)
, \\
D_{+-}^{\tau\nu}({\bf q}) =&  -\sum_{{j}} ({\bf q} + {\bf
G}_j)_{\tau}({\bf q} + {\bf G}_j)_{\nu} e^{i {\bf G}_j \cdot {\bf
c}} \Upsilon \Big( \tfrac{|{\bf G}_j + {\bf q}|}{2 \alpha},\alpha l \Big)\nonumber\\& + \sum_{{j}} e^{-i {\bf q} \cdot ({\bf R}_j + {\bf
c})} \lim_{{\bf r} \rightarrow 0}
\partial_{\tau}\partial_{\nu} \Omega_2 \big(|\tilde{\bf R}_j + {\bf r}|\big) ,
\end{align}\label{eq:eqRap}\end{subequations}where $\delta_{\tau\nu}$ is the Kronecker-delta, and where we used the
notation $({\bf q}+{\bf G}_j)_\tau\equiv {\bf e}_\tau\cdot({\bf
q}+{\bf G}_j)$ for the component in the direction $\tau$ of the
two-dimensional (reciprocal) vectors. The functions
$\Upsilon(x, y)$, $\Omega_1(x)$ and $\Omega_2(x)$ in Eq.~\eqref{eq:eqRap} are given
by
\begin{subequations}
\begin{align}
\Upsilon (x, y) =& \frac{4 \alpha}{\sqrt{\pi}} e^{-x^2
-y^2}+\sum_\pm \left(\pm 2\right) \alpha x e^{\pm 2 x y} {\rm erfc} \left( x \pm
y
\right),\\
\Omega_1(x) =& \frac{\mathrm{erfc} (\alpha x)}{x^3} + \frac{2
\alpha
~e^{-\alpha^2 x^2}}{\sqrt{\pi} x^2} , \\
\Omega_2(x) =&~ \frac{\mathrm{erfc} (\alpha x)}{x^3} + \frac{2
\alpha
~e^{-\alpha^2 x^2}}{\sqrt{\pi} x^2} \nonumber\\& -3 l^2 \Bigg[ \frac{\mathrm{erfc}
(\alpha x)}{x^5} + \frac{2 \alpha ~(3 + 2 \alpha^2 x^2) ~e^{-\alpha^2 x^2}}{3
\sqrt{\pi} ~x^4} \Bigg].
\end{align}
\end{subequations}

For the configurations of Fig.~\ref{fig:fig01:lattice}, the complex Hermitian matrix ${\bf M}_{\rm D}({\bf q})$ of Eq.~\eqref{eq:7} can be transformed into a real and symmetric matrix. This is achieved by first applying the unitary transformation $\bar{\bf M}_{\rm D}({\bf q})={\bf U}{\bf M}_{\rm D}({\bf q}){\bf U}^\dag$, with ${\bf U}$ a $4 \times 4$ matrix defined as
\begin{align}
{\rm \bf U} =& \frac{1}{\sqrt{2}}\left(
\begin{array}{ll}
~{\rm \bf I}_2 & i {\rm \bf I}_2  \\
i {\rm \bf I}_2 & ~{\rm \bf I}_2
\end{array}\right),
\end{align}
with ${\rm \bf I}_2$ the $2 \times 2$ identity matrix. This transformation brings the dynamical matrix into the symmetric form
\begin{subequations}
\begin{align}
\bar{\bf M}_{\rm D}({\bf q})=& \left(
\begin{array}{cc}
{\rm \bf M}_{++}+ {\rm Im}[{\rm \bf M}_{+-}] &  {\rm Re}[{\rm \bf M}_{+-}]  \\
{\rm Re}[{\rm \bf M}_{+-}] & {\rm \bf M}_{++}- {\rm Im}[{\rm \bf
M}_{+-}]
\end{array}\right),\\
&\mbox{with\quad }
 {\rm \bf M}_{+\sigma} \equiv \left(
\begin{array}{ll}
M_{+\sigma}^{xx} & M_{+\sigma}^{xy}  \\
M_{+\sigma}^{xy} & M_{+\sigma}^{yy}
\end{array}
\right).
\end{align}
\end{subequations}
That the matrix $\bar{\bf M}_{\rm D}({\bf q})$ is real now stems from the fact that ${\rm Im}[{\rm \bf M}_{+-}]$ vanishes for a lattice with inversion symmetry~\cite{goldoni,wallace}, which is the case for all lattice configurations considered in this work.

For each quasi-momentum ${\bf q}$, diagonalizing $\bar{\bf M}_{\rm
D}({\bf q})$ provides the square of the phonon frequencies,
$\omega_\alpha({\bf q})^2$ (with $1\leq  \alpha\leq 4$), which
correspond to the four distinct phonon modes of the bilayer system.
Within the harmonic approximation, the stability of the various
lattice configurations is linked to the sign of the eigenvalues of
$\bar{\bf M}_{\rm D}({\bf q})$. That is, {a lattice configuration is
stable if all four eigenvalues of $\bar{\bf M}_{\rm D}({\bf q})$ are
positive for all quasi-momenta ${\bf q}$, i.e. $\omega_\alpha({\bf
q})^2>0$, while it is unstable if one (or more) of the four
eigenvalues is negative for a given quasi-momentum ${\bf q}$, c.f.
$\omega_\alpha({\bf q})^2< 0$}.

%----------------------------------------------------------

%%% GP

%----------------------------------------------------------

\subsection{Ground state configuration}\label{dyna:ground}

\begin{figure}
\includegraphics[width=\columnwidth]{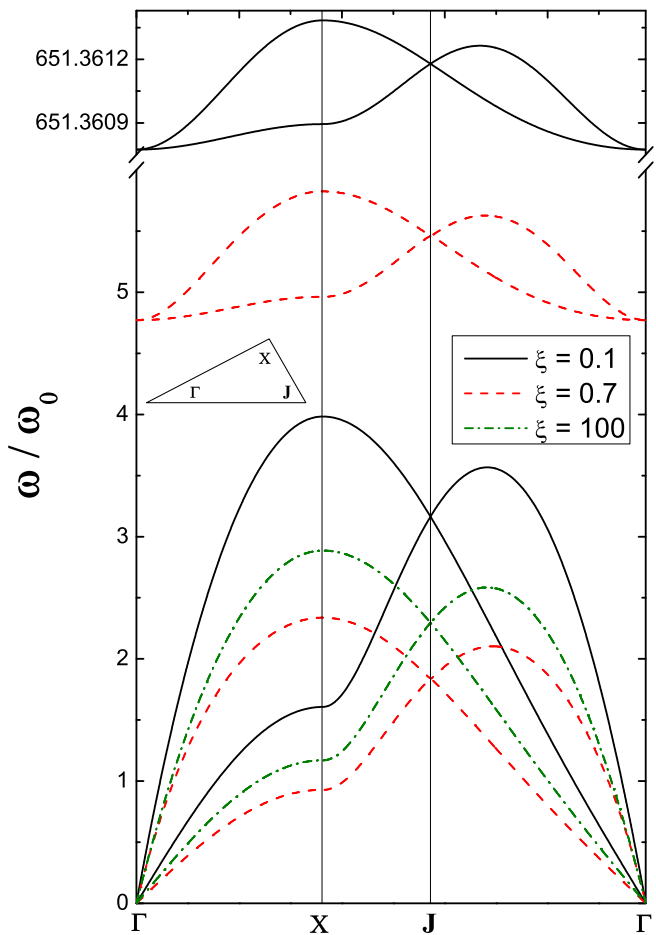}
\caption{(Color online) Phonon dispersion curves {
$\omega_{\alpha}({\bf q})$} for the MH lattice configuration in
units of $\omega_0 = \sqrt{d^2 n^{5/2}/m}$, {for $\xi=0.1$ (solid
lines), $\xi=0.7$ (dashed lines) and $\xi\rightarrow\infty$
(dash-dotted lines)}. The frequencies are presented along the high
symmetry directions in the Brillouin zone. The high-symmetry points
$\Gamma$, $\mathrm{X}$ and $\mathrm{J}$ are {depicted} in the inset.
{Note that the different axis scaling for the high-frequency regime
$\omega\gg6\omega_0$.}} \label{fig:fig04:phmh}
\end{figure}

\begin{figure}
\includegraphics[width=\columnwidth]{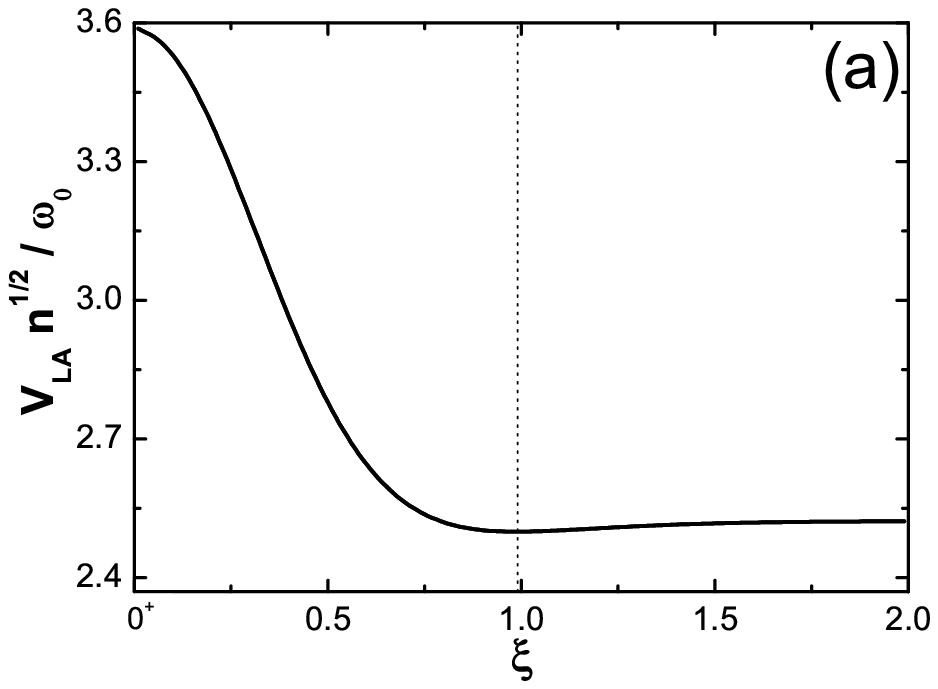}
\includegraphics[width=\columnwidth]{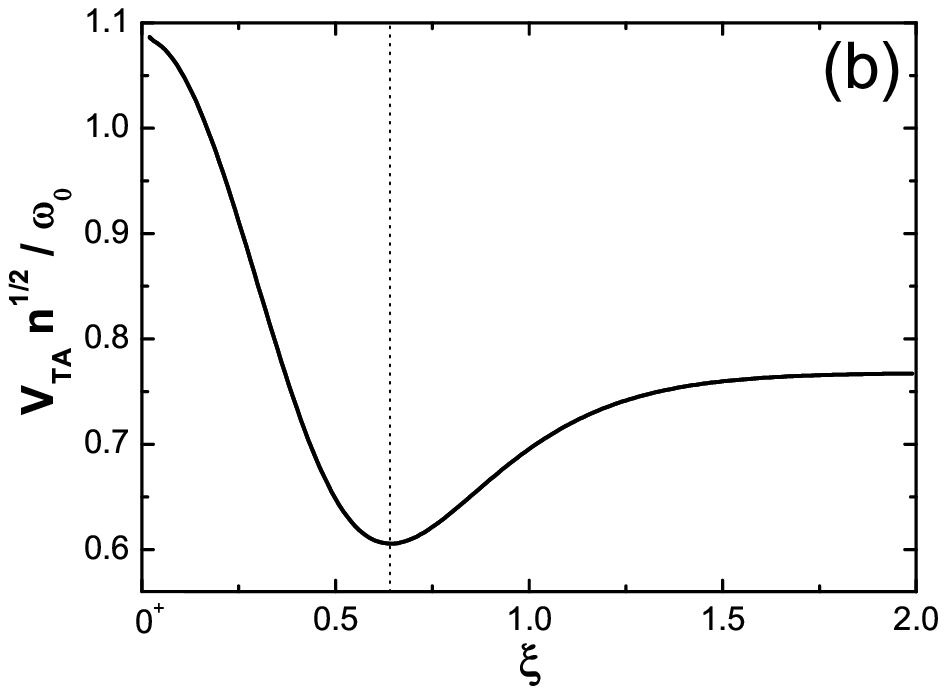}
\caption{\label{fig:fig05:svmh}(Color online) {Sound velocities for
the MH lattice configuration as a function of the dimensionless
layer separation $\xi$: (a) Longitudinal sound velocity $v_{\rm LA}$
and (b) transverse sound velocity $v_{\rm TA}$, in units of
$\omega_0/\sqrt{n}$.}}
\end{figure}

\begin{figure}
\includegraphics[width=\columnwidth]{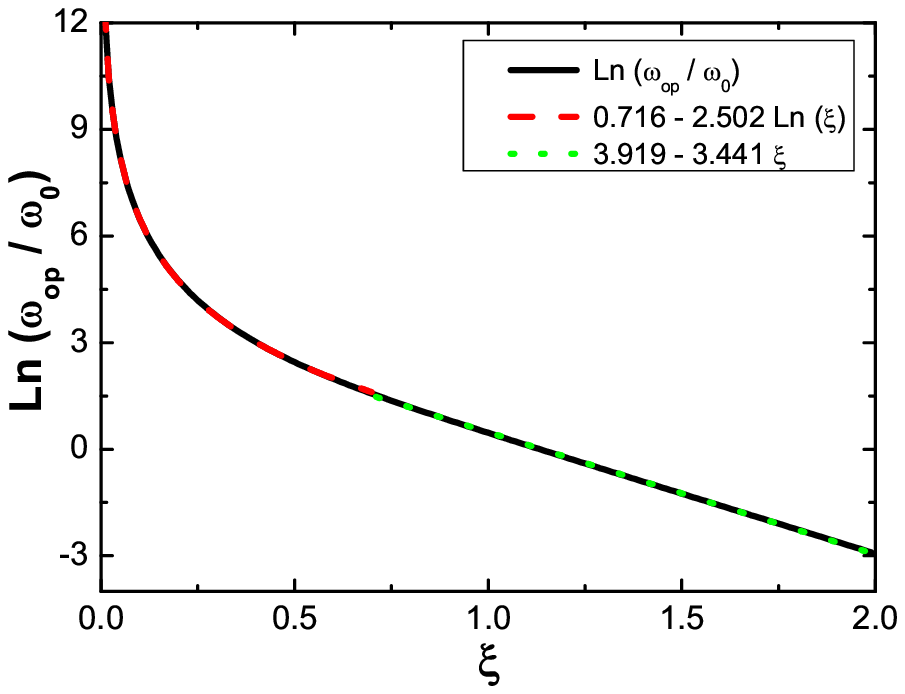}
\caption{(Color online) {Optical frequencies $\omega_{\rm op}$ (in
logarithmic scale in units of $\omega_0$) at the $\Gamma$ point for
the MH lattice configuration as a function of $\xi$ (solid line),
along with the approximations $\ln(\omega_{\rm op}/\omega_0)\approx
0.716-2.502\ln\xi$ for $\xi \lesssim 0.7$ (dashed line) and
$\ln(\omega_{\rm op}/\omega_0) \approx 3.919-3.441 \xi$ for
$\xi\gtrsim0.7$ (dotted line).}}
 \label{fig:fig06:woc0}
\end{figure}

In this subsection we calculate the phonon spectrum for the
ground-state configuration MH as a function of the inter-layer
separation $\xi$, using the techniques described above. As expected,
we find both acoustic and optical modes, with the latter related to
out-of-phase vibrations of particles in different layers. We show
that the longitudinal and transverse sound velocities of the
acoustic modes show a non-monotonic dependence on $\xi$, which {is
linked} to the anisotropic nature of the dipole-dipole interaction.
The picture emerging from these studies is one where the bilayer
structure behaves for vanishing inter-layer separations $\xi \ll 1$
as a single 2D crystal of particles with double mass and double
dipole-moment, while for $\xi\gg1$ the two layers behave as
independent {2D crystals}. Within the harmonic approximation
inherent {to} the present discussion, the transition between these
two situations occurs approximately for inter-layer distances such
that the nearest-neighbor inter-layer interaction switches from
repulsive
($\xi\ll 1$) to attractive ($\xi \gg 1$).\\

Figure~\ref{fig:fig04:phmh} shows the phonon dispersion relations
for the MH lattice configuration {along the} high symmetry
directions in the Brillouin zone for a {three values of $\xi$, i.e.
$\xi=0.1,0.7,100$ (solid, dashed, dash-dotted lines)}. The
frequencies are in units of the characteristic phonon frequency
$\omega_0=\sqrt{d^2 n^{5/2}/m}$. %The phonon spectra are shown along the
%characteristic symmetry directions in reciprocal space.
The symmetry points ${\rm \Gamma}$, ${\rm X}$ and ${\rm J}$ are
depicted in the inset. The figure shows that for $\xi \rightarrow
\infty$ there are two phonon branches. Each one of these branches is
doubly degenerate, corresponding to independent longitudinal and
transverse acoustic phonon modes of the two layers. We find that, as
expected, the modes exactly match those of a single layer,
confirming that in the limit $\xi \rightarrow \infty$ the two layers
behave as independent. For {finite $\xi$} the dipole-dipole
interaction couples the two layers, and lattice vibrations in the
two layers become correlated. Accordingly, the figure shows that for
$\xi=0.7$ {(dashed lines)} and {$\xi=0.1$ (solid lines)} the phonon
modes develop separate acoustic and optical branches. Since the MH
lattice structure can be represented as the repetition of a basis
cell comprising two particles, one per layer, stacked on top of each
other, the optical modes are easily understood as arising from
out-of-phase vibrations of the dipoles in each basis cell. The
figure shows that the optical frequency of vibration increases with
decreasing inter-layer distance, and the bandwidth of the optical
branch tends to shrink (note that in the figure the scales for the
optical modes for $\xi=0.7$ and {$\xi=0.1$} are different). In the
(unphysical) limit $\xi\rightarrow 0$, the phonon spectrum reduces
to one where there is only one, i.e. {\it non-degenerate},
longitudinal acoustic and one transverse acoustic mode, while the
optical branches are shifted to infinitely large frequencies. This
observation is consistent with the system behaving as a single layer
of dipoles with double mass and double dipole strength, arranged in
a hexagonal configuration (see below).

Figure~\ref{fig:fig04:phmh} shows that the acoustic branches for
$\xi=0.7$ have lower frequencies than the corresponding branches for
$\xi=0.1$ and $\xi \rightarrow \infty$, indicating a non-monotonic
dependence of the frequencies on $\xi$. In order to better
investigate this point, in {Fig.~\ref{fig:fig05:svmh}(a) and
Fig.~\ref{fig:fig05:svmh}(b)} we plot the longitudinal and
transverse sound velocities $v_{\textrm{LA}} =
\partial \omega_{\textrm{LA}}/\partial q ~|_{q = 0}$, $v_{\textrm{TA}} =
\partial \omega_{\textrm{TA}}/\partial q ~|_{q = 0} $, with
 $\omega_{\textrm{LA}}$ and $\omega_{\textrm{TA}}$ the frequencies {of
the} longitudinal and transverse acoustic modes, respectively. We
find that for $\xi \gg 1$, the longitudinal and transverse sound
velocities tend to $v_{\textrm{LA}} \simeq 2.544 \omega_0/ \sqrt{n}$
and $v_{\textrm{TA}} \simeq 0.768 \omega_0/ \sqrt{n}$, respectively.
These values correspond to the sound velocities of a monolayer
hexagonal lattice configuration, consistent with the observation
above that for $\xi\gg 1$ the crystal vibrations in the two layers
become independent. In the opposite limit $\xi \rightarrow 0 $, we
find that the sound velocities are larger than those above {\it
exactly} by a factor $\sqrt{2}$. A simple comparison with the
characteristic phonon frequency $\omega_0=\sqrt{d^2n^{5/2}/m}$
shows that this $\sqrt{2}$-factor is consistent with the system
behaving as a single layer of dipoles of mass $2 m$ and dipole
strength $2 d$.

{Figure~\ref{fig:fig05:svmh}} shows that the dependence of the sound
velocities on $\xi$ is non-monotonic. In particular, minima for
$v_{\rm LA}$ and $v_{\rm TA}$ occur at $\xi = \xi_0 \approx 1$ and
0.64, respectively {(see Appendix~\ref{sec:appB})}. This
non-monotonic behavior is {linked} to the anisotropic character of
the dipole-dipole interaction. In fact, for $\xi \rightarrow \infty$
the two layers behave as independent hexagonal crystals. For finite
$\xi$, the inter-layer interactions couple the two layers. This
coupling, which is responsible for the formation of the optical
band, splits the degeneracy of the phonon frequencies, lowering the
acoustic sound velocity, which is associated with a softening of the
crystal, see Fig.4. For $\xi \lesssim 1$ the optical and acoustic
branches are well separated, corresponding to the formation of a
crystal of tightly bound pairs of dipoles. For $\xi \rightarrow 0$
the sound velocity is a $\sqrt{2}$ larger than at $\xi \rightarrow
\infty$, as discussed above, and this determines the appearance of a
minimum in between, and we observe that the value $\xi \approx 1$
roughly corresponds to the inter-layer distance at which the
interaction of a dipole at ${\bf R}_{+,j}$ with
next-nearest-neighbor in the opposite layer at position ${\bf
R}_{-,j'}$ switches from attractive to repulsive. We remark that
although the crystalline structure is softened for finite $\xi$, the
sound velocity remains always finite. This is due to the anisotropic
character of the dipole-dipole interactions, which ensures that the
MH configuration is {\em always} the {\em stable} ground-state
configuration.

Figure~\ref{fig:fig06:woc0} shows {as a solid line} the optical
frequencies $\omega_{\rm op}$ as a function of $\xi$ in logarithmic
scale at the $\Gamma$ point, where the two optical frequencies are
degenerate. We notice that for $\xi \gtrsim 0.7$ the frequencies
decay exponentially with increasing $\xi$ as {$\ln(\omega_{\rm
op}/\omega_0) \approx 3.919-3.441 \xi$ (c.f. dotted line), while
they diverge as a power law $\ln(\omega_{\rm op}/\omega_0) \approx
0.716-2.502 \ln(\xi)$ for $\xi \lesssim 0.7$ (c.f. dashed line)}.
This change in behavior is another manifestation of
the crossover from two independent layers to a single crystal of paired dipoles.%
%
%

%%% GP

%

%
%

%

%

%%%%%%%%%%%%%%%%%%%%%%%%%%%%%%%%%%%%%%%%%%%%%%%%%%%%%%%%%%%%%%%%%%

%%% GP start

%----------------------------------------------------------

%\subsection{Meta-stable configurations}\label{dyna:meta}
\subsection{Excited-state configurations}\label{dyna:meta}

\begin{figure}
\includegraphics[width=\columnwidth]{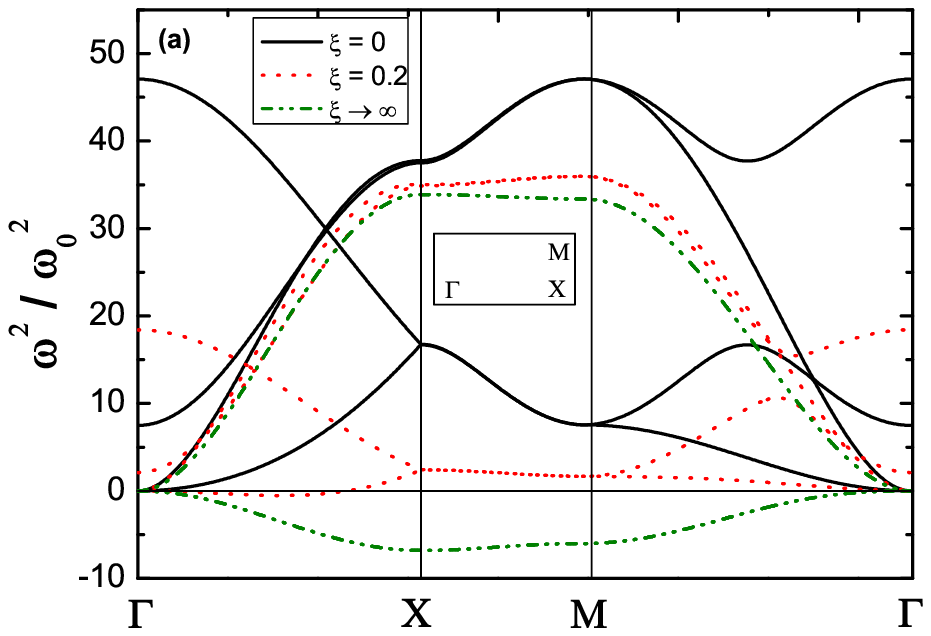}
\includegraphics[width=\columnwidth]{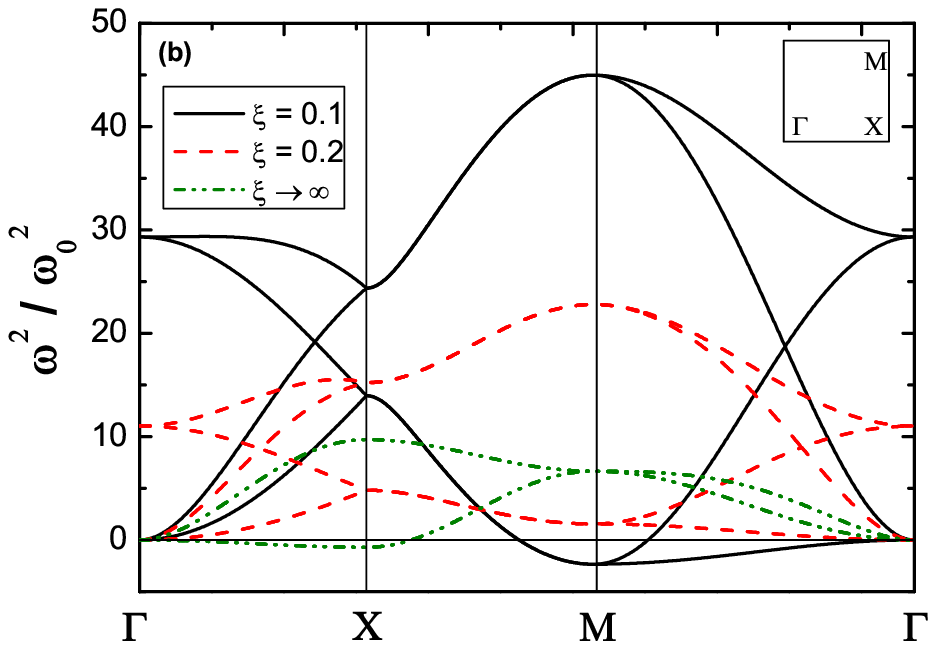}
\caption{(Color online) Square of the phonon frequencies
$\omega_\alpha^2$ {in units of $\omega_0^2 = d^2 n^{5/2}/m$ (a) for
the OCH configuration {at $\xi=0$ (solid lines), $\xi=0.2$ (dashed
lines) and $\xi\rightarrow\infty$ (dash-dotted lines)} and (b) for
the ZS configurations {at $\xi=0.1$ (solid lines), $\xi=0.2$ (dashed
lines) and $\xi\rightarrow\infty$ (dash-dotted lines)}. The
frequencies are {shown} along the high symmetry directions in
reciprocal space for each lattice configuration. The high-symmetry
points $\Gamma$, $\mathrm{X}$ and $\mathrm{M}$ are {depicted} in the
insets.}} \label{fig:fig07:phsoft}
\end{figure}

\begin{figure}
\includegraphics[width=\columnwidth]{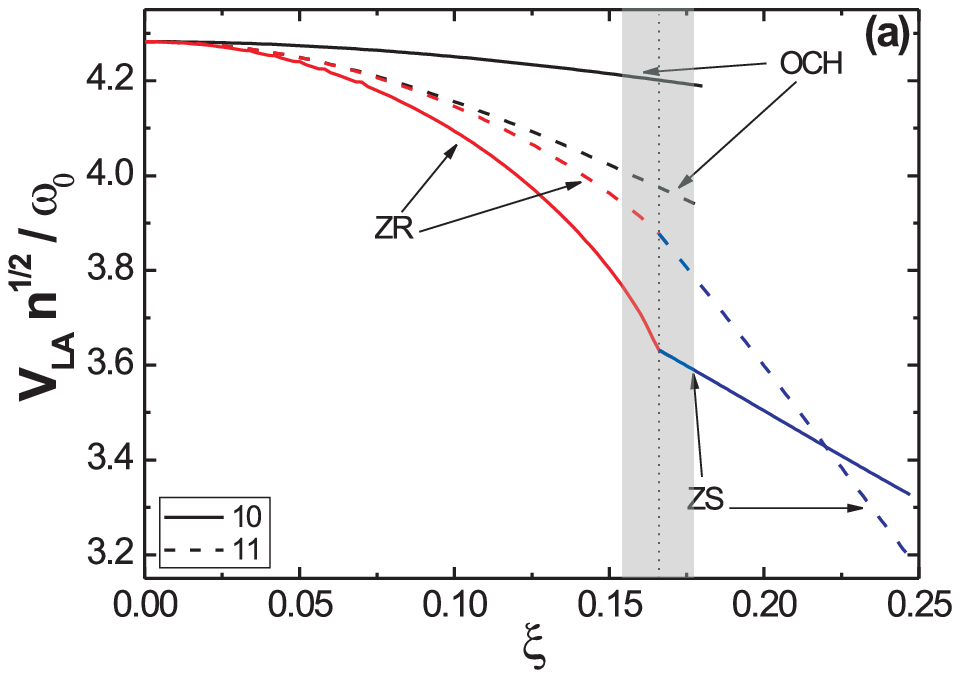}
\includegraphics[width=\columnwidth]{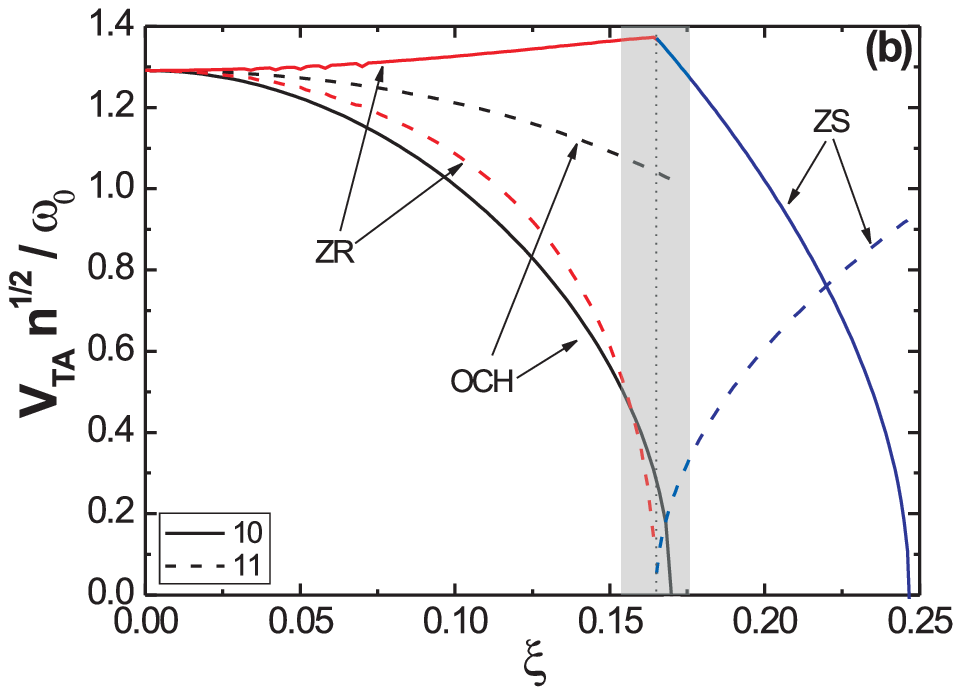}
\caption{(Color online) {(a) Longitudinal sound velocities $v_{\rm
LA}$ and (b) transverse sound velocities $v_{\rm TA}$} in units of
$\omega_0/\sqrt{n}$ along the (1,0) (solid lines) and the (1,1)
(dotted lines) directions for OCH, ZR and ZS lattice configurations.
The vertical dotted line denotes the boundary between ZR and ZS.
{The gray shading corresponds to the transition region where the
harmonic approximation for the phonon excitations may become
inadequate.}}
\label{fig:fig08:svsoft}
\end{figure}

In this subsection we study the {\em meta}-stability of the
excited-state configurations OCH, ZS {and ZR} introduced in Sect.II,
by calculating the phonon modes for each configuration. For a given
interlayer distance $\xi$, within the applicability of the harmonic
approximation, regions of meta-stability and instability for the
above configurations correspond to real and imaginary values of the
computed phonon frequencies, respectively. By analyzing the sound
velocities of the phonon excitations, we find that the instability
of the OCH and ZS crystalline structures is associated with the
vanishing of the transverse acoustic branch of the phonon modes in
the directions $\Gamma \textrm{X} $ and $\Gamma \textrm{M}$. We thus
derive a stability diagram for the low-lying excitations of the
system as a function of $\xi$, see below. This is interesting, since
in principle these excited-state configurations may be realized,
e.g. by first trapping cold polar molecules in optical lattices with
the same geometry as OCH and ZS crystals, increasing the
dipole-dipole interactions using external fields, forming
interaction-induced OCH and ZS crystals, and finally
adiabatically removing the lattice potential.\\

Panels (a) and (b) in Fig.~\ref{fig:fig07:phsoft} show the square of
the phonon frequencies for the OCH and ZS lattice configurations,
respectively, for a few values of $\xi$ and in units of $\omega_0^2
= d^2 n^{5/2}/m$. The phonon spectra in Fig.~\ref{fig:fig07:phsoft}
are shown along the high symmetry directions in reciprocal space,
with symmetry points labeled in the insets. The figure shows that
for certain values of $\xi$ the square of the
phonon frequencies becomes negative, $\omega^2 < 0$, %. This is
%interesting, since negative values of $\omega^2$
signaling an instability of the corresponding crystalline structure
for the given value of $\xi$. In panels (a) and (b), regions of
meta-stability are present for $\xi=0$ and {$\xi=0.2$},
respectively. In fact, we determined numerically that the OCH
configuration of panel (a) is {\it meta}-stable for $\xi \lesssim
0.170$, while the {ZS configuration} of panel (b) is {\it
meta}-stable in the range $0.166 \lesssim \xi \lesssim 0.247$. We
notice that in these regimes where $\omega^2>0$ the system in the
OCH and ZS configurations is {\it meta}-stable, since the associated
crystalline structures are excited states of the system. By
analyzing the sound velocities of the phonon excitations, we found
that the instability of the OCH and ZS crystalline structures is
associated with the vanishing of the transverse acoustic branch of
the phonon modes in the directions $(1,1)$ and $ (1,0)$, as detailed
below.\\

Panels (a) and (b) in Fig.~\ref{fig:fig08:svsoft} show the sound
velocities of the transverse acoustic (TA) and longitudinal acoustic
(LA) modes for the OCH, ZS and ZR configurations as functions of
$\xi$ in the range of stability of each configuration, respectively.
In the figure, the continuous and dashed lines correspond to the
(1,0) and (1,1) directions, respectively. Panel (a) shows that the
TA mode for the OCH configuration vanishes in the (1,0) direction at
$\xi\simeq 0.170$ [see also Fig.~\ref{fig:fig07:phsoft}(a)]. The TA
mode for the ZS configuration vanishes in the (1,1) and (1,0)
directions for $\xi\lesssim 0.166$ and $\xi \gtrsim 0.25$,
respectively. In addition, panel (b) shows that the longitudinal
modes for all three configurations soften with increasing $\xi$. The
transverse and longitudinal sound velocities for the ZR
configuration interpolate between those of the OCH and ZS
configurations, {see Fig.~\ref{fig:fig08:svsoft}(a) and
Fig.~\ref{fig:fig08:svsoft}(b)}, coinciding with those at $\xi=0$
and $\xi\approx0.166$, respectively.

\begin{figure}[ht!]
\includegraphics[width=\columnwidth]{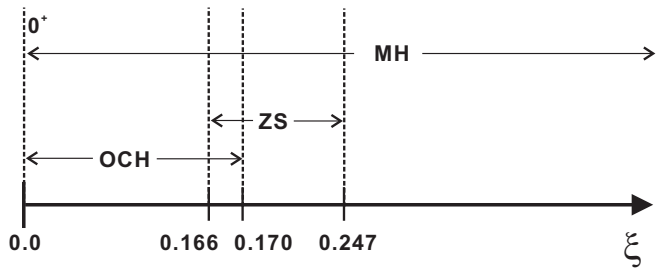}
\caption{Approximate regions of stability for the MH, ZS, and OCH
configurations, as a function of $\xi$.} \label{fig:fig10:phase}
\end{figure}

Figure~\ref{fig:fig10:phase} summarizes the regions of stability for
the MH configuration, which is the ground-state for any $\xi>0$, the
OCH and the ZS configurations . We remark that the values $\xi =
$0.166, 0.170 and 0.247 have been {obtained numerically}, in the
harmonic approximation for the phonon spectrum. Since the harmonic
approximation is bound to break down around any transition points
between various configurations, the numerical values 0.166 and 0.170
should be taken with caution, and simply interpreted as {\it
indicative} of the transition region.

\begin{figure}[b!]
\includegraphics[width=\columnwidth]{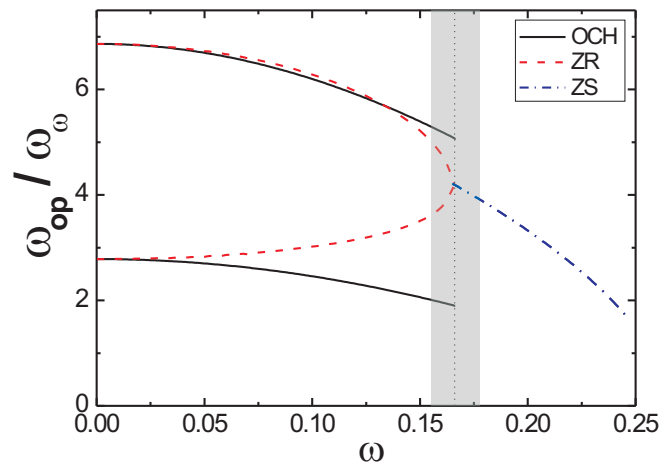}
\caption{\label{fig:fig09:wosoft}(Color online) Optical frequencies
$\omega_{{\rm op}}$ {in unit of $\omega_0$} at the $\Gamma$ point
for the OCH, ZR and ZS lattice configurations. The vertical dotted
line denotes the boundary between ZR and ZS. {The gray shading
corresponds to the transition region where the harmonic
approximation for the phonon excitations may become inadequate.}}
\end{figure}

For completeness, in Fig. \ref{fig:fig09:wosoft} we show the optical
frequencies at the $\Gamma$ point as function of $\xi$ for the OCH,
ZR and ZS lattice configurations. As expected, we find that the
frequencies corresponding to the ZR configuration interpolate
between those of the OCH and ZS configurations. In particular, they
become degenerate around $\xi \approx 0.170$, where the ZS
configuration becomes stable. This degeneracy is a natural
consequence of the fact that the aspect ratio $a_2/a_1$ of the ZS
configuration is 1. We notice that the different behaviors of the
optical frequencies may be used to distinguish experimentally the
various metastable structures, when initially prepared in tailored
optical lattice potentials, as discussed at the beginning of this
subsection.

\section{classical melting}\label{sec:sec4}

In this section we discuss the classical melting temperature of the
bilayer system, as obtained in the harmonic approximation for the excitations of the crystal using a (modified) Lindemann criterion. The latter states that for a given configuration the melting occurs when the {\em mean relative displacement} between neighboring sites becomes of the order of the mean interparticle distance $r_0 = 1/\sqrt{\pi n}$ (see Table.~\ref{tab:tab1} for the configurations of Fig.~\ref{fig:fig01:lattice})\cite{bedanov}. This reads
\begin{eqnarray}
\frac{\sqrt{\langle [{\bf u}({\bf R})- {\bf u}({\bf R}+ {\bf r})]^2 \rangle}}{r_0} = \delta_m,\label{eq:eqLindemann}
\end{eqnarray}
where $\langle |{\bf u}({\bf R})- {\bf u}({\bf R}+ {\bf r})|^2 \rangle$ is the relative
mean square displacement, ${\bf u}({\bf R})$ and ${\bf u}({\bf R}+
{\bf r})$ are the displacement vectors at site ${\bf R}$ and at its nearest
neighbor site ${\bf R} + {\bf r}$, $\langle \rangle$ is a thermal average, and $\delta_m<1$ is a parameter which in general has to be determined numerically.

The left-hand side of Eq.~\eqref{eq:eqLindemann} is computed in the
harmonic approximation for the phonons as follows. Each particle in
the bilayer structure has two different kinds of nearest-neighbors:
in-plane and out-of-plane. Thus, in analogy to the Coulomb case of
Ref.~\cite{goldoni} we define the {\em intralayer} ($\Delta u_{++}$)
and {\em interlayer} ($\Delta u_{+-}$) correlation functions {(see
Appendix~\ref{sec:appC})}
\begin{subequations}
\begin{eqnarray}
\Delta u_{++} &=& \frac{1}{S_+} \sum_{\gamma=x,y} \sum_{h=1\ldots
S_+}\langle |u_{\gamma}^+(0) - u_{\gamma}^+(h)|^2 \rangle \nonumber
\\
&=& \frac{2 k_B T}{N S_+ m} \sum_{{\bf q},j} \frac{1}{\omega^2({\bf
q},j)} \Bigg\{ \left[ \rho_{x}^+({\bf q},j)^2
+ \rho_{y}^+({\bf q},j)^2 \right] \nonumber\\
&&\left[ 1 - \cos ({\bf q} \cdot {\bf R}_h^+) \right] \Bigg\} , \label{eq:cf1}\\
%
%\end{eqnarray}
%
%\begin{eqnarray}
\Delta u_{+-} &=& \frac{1}{S_-} \sum_{\gamma=x,y} \sum_{h=1\ldots
S_-}\langle |u_{\gamma}^+(0) - u_{\gamma}^-(h)|^2 \rangle \nonumber
\\
&=& \frac{k_B T}{N S_- m} \sum_{{\bf q},j} \frac{1}{\omega^2({\bf
q},j)} \Bigg\{ \Big[ \rho_{x}^+({\bf q},j)^2
+ \rho_{x}^-({\bf q},j)^2 \nonumber\\
&& + \rho_{y}^+({\bf q},j)^2 + \rho_{y}^-({\bf q},j)^2 \Big] - 2
\Big[ \rho_{x}^+({\bf q},j) \rho_{x}^-({\bf q},j) \nonumber\\
&& + \rho_{y}^+({\bf q},j) \rho_{y}^-({\bf q},j) \Big] \cos\left(
{\bf q} \cdot {\bf R}_h^- \right) \Bigg\}. \label{eq:cf2}
\end{eqnarray}
\end{subequations}

\begin{figure}
\includegraphics[width=.9\columnwidth]{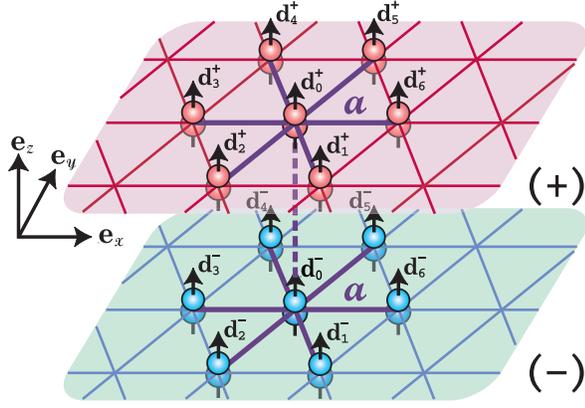}
\caption{(Color online) {Illustration of the dipole-configuration
for the MH structure leading to the modified Lindemann
criterion}.}\label{fig:figConfiguration}
\end{figure}

Here {$S_\sigma$} is the number of nearest-neighbor {dipoles} in
layer {$\sigma=\pm$, $u_{\gamma}^{\sigma}(h)$} is the {$\gamma^{\rm
th}$} component of the displacement of a particle at position $h$ in
the layer {$\sigma$}, {$\rho_{\gamma}^{\sigma}({\bf q},j)$} is the
{$\gamma^{\rm th}$} component of the eigenvector of the {$j^{\rm
th}$} mode at point ${\bf q}$ in the Brillouin zone of the
sublattice in layer {$\sigma$}, and {${\bf R}_h^{\tau}$} is the
relative vector connecting one particle to its {$h^{\rm th}$}
nearest-neighbor in the same {($\tau=+$) or opposite ($\tau=-$)}
layers. We notice that the number of nearest-neighbors {$S_\sigma$}
and their distance depends on the considered lattice configuration
(see Fig.~\ref{fig:fig01:lattice}) and $\xi$. In particular,
Fig.~\ref{fig:figConfiguration} shows that, in the relevant case of
the groundstate configuration MH (see Sect.~\ref{sec:sec4a} below),
the number of {\em in-plane} nearest-neighbors is 6, while that of
{\em out-of-plane} nearest-neighbors is 7.

The correlation $\langle [{\bf u}({\bf R})- {\bf u}({\bf R}+ {\bf r}) ]^2 \rangle$ of Eq.~\eqref{eq:eqLindemann} is now computed as
\begin{eqnarray}
\langle [{\bf u}({\bf R})- {\bf u}({\bf R}+ {\bf r})]^2 \rangle = \Delta u_{++} + f(l)\Delta u_{+-},
\end{eqnarray}
where the function $f(l)$ describes the influence
of lattice vibrations in one layer on the lattice vibrations in the opposite layer, and it is defined as
\begin{eqnarray}
f(l) = \frac{1}{(1 + \kappa l^2)^{5/2}} + \frac{-3 \kappa l^2}{(1 +
\kappa l^2)^{7/2}}.\label{eq:eqFL}
\end{eqnarray}
Here, the geometric parameter $\kappa$ can be obtained from
Table~\ref{tab:tab1}, and it reads $\kappa = (a^2)^{-1}$ and $\kappa
= (|{\bf c}|^2)^{-1}$ for the MH and zigzag lattice configurations,
respectively. This expression for $f(l)$ is chosen to be
proportional to the in-plane part of the force between two
nearest-neighbor {dipoles} in opposite layers and it satisfies the
conditions
\[ \lim_{l\rightarrow 0} f(l)= 1 {\quad\mbox{and}\quad} \lim_{l\rightarrow \infty}f(l) = 0,\]
where the latter condition is due to the fact that vibrations in the
two layers are independent for infinite interlayer separations.

%----------------------------------------------------------

%

%
%%----------------------------------------------------------

\subsection{Melting of the ground state configuration}\label{sec:sec4a}

In this subsection we determine the classical melting temperature
{$T_{\rm m}$} of the ground-state crystal configuration MH as a
function of $\xi$, using the modified Lindemann criterion introduced
above. We find a non-monotonic dependence of {$T_{\rm m}$} on $\xi$,
which we attribute to the anisotropic nature of the dipole-dipole
interactions. This is interesting since for certain temperatures it
is associated with a {\em re-entrant} melting
behavior in the form of {\it solid-liquid-solid-liquid} transitions.\\

\begin{figure*}[htbp!]
\includegraphics[width=\textwidth]{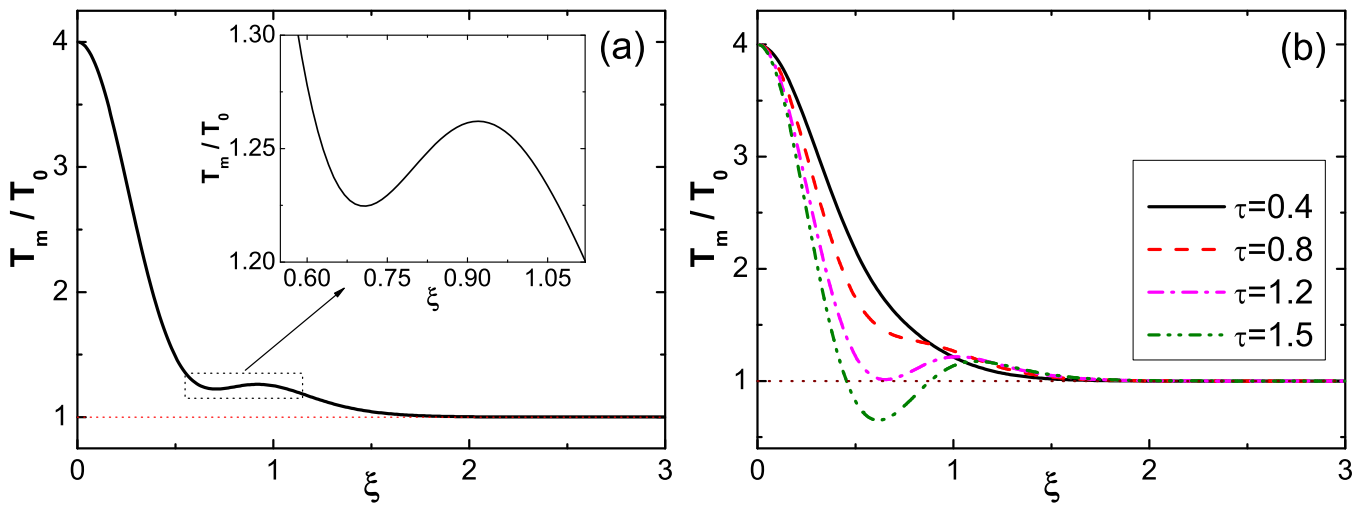}
\caption{(Color online) (a) Melting temperature {$T_{\rm m}$} as a
function of $\xi$ for the MH lattice configuration {(solid line)}.
The unit $T_0 \approx 0.066 d^2/a^3$ corresponds to the melting
temperature of a single-layer crystal, as computed from
Eq.~\eqref{eq:eqLindemann} with $\delta_{\rm m}=0.23$. {The dashed
line is a guide to the eye}. (b) Melting temperature {$T_m$} as a
function of $\xi$ for particles interacting via the potential
$V_{ij}^{\mathrm{int}}$ of Eq.~\eqref{eq:eqVtunable}. Here, $\tau$
is the strength of the attractive part of the potential (see text),
with $\tau=1$ corresponding to the dipole-dipole interaction of
panel (a). {The dotted line is a guide to the eye.}}
\label{fig:fig11:tmmh}
\end{figure*}

Figure~\ref{fig:fig11:tmmh} shows the melting temperature $T_{\rm m}$ as a
function of $\xi$, as calculated from the Lindemann criterion with
$\delta_{\rm m}=0.23$ within the harmonic approximation for the phonon
modes. The precise value of $\delta_{\rm m}$ should in principle be obtained numerically, e.g. using molecular dynamics simulations. In the absence of such a computation for a classical bilayer crystal, the value of $\delta_{\rm m}=0.23$ has been chosen in analogy to the one obtained in Ref.~\cite{Astrakharchik07} for the {\em quantum} melting
transition from a single-layer crystal of bosonic dipoles into a superfluid using Diffusion Monte-Carlo techniques. %Since it has been obtained numerically, we notice that this value of $\delta_{\rm m}$ has not been derived in the harmonic approximation for the crystal excitations.
Using this value of $\delta_{\rm m}$ in Eq.~\eqref{eq:eqLindemann},
we find that for $\xi\gg1$ the classical melting temperature of the
bilayer crystal tends to the value $T_0\approx 0.066 d^2/a^3$. By
construction, the latter corresponds to the classical melting
temperature of a single hexagonal crystal as computed in the
harmonic approximation discussed above. We here notice that the
obtained value of {$T_0$ is of the order of the actual one $T_0^{\rm
num}\approx 0.089 d^2/a^3$} for a classical single-layer crystal, as
obtained numerically by molecular dynamics simulations \cite{Kalia}.
Since the spirit of the Lindemann criterion is that of a {\it
qualitative} estimate of the transition point, in the following we
will be content with the value $T_0$.

For $\xi \ll 1$ the figure shows that the melting temperature tends
to $T_{\rm m} = 4 T_0$. This is consistent with the picture of a
hexagonal crystal made of paired dipoles with dipole strength
$d^{\star}= 2d$, as discussed in {Sect.~\ref{dyna:ground}}.
Interestingly, the figure shows that $T_m$ has a non-monotonic
dependence on $\xi$ around $\xi\approx 1$. In particular, the
$T_{\rm m}$-vs-$\xi$ curve has a local maximum and a local minimum
at $\xi\approx1$ and $\xi\approx0.6$, respectively. As noticed in
{Sect.~\ref{dyna:ground}}, this region of $\xi$-values corresponds
to the distance at which the dipole-dipole interaction between a
particle in one layer and its nearest-neighbor in the opposite layer
changes sign from attractive to repulsive (e.g. $d_0^-$ and $d_j^+$
in Fig.~\ref{fig:fig01:lattice}, respectively, with $1\leq j\leq
6$){.}

In order to check that the non-monotonicity is in fact connected with the anisotropic nature of the dipole-dipole interaction, in Fig~\ref{fig:fig11:tmmh}(b) we have plotted the melting temperature for an {\em artificial} system of particles where the strength of the attractive part of the dipole-dipole interaction can be tuned, and thus the particles interact via a potential of the form
\begin{eqnarray}
V_{ij}^{\mathrm{int}} = d^2 \left( \frac{1}{|{\bf r}_{ij}|^3} + \tau
~\frac{-3 ~l^2}{|{\bf r}_{ij}|^5} \right),\label{eq:eqVtunable}
\end{eqnarray}
with $\tau$ a constant, $0\leq\tau<\infty$, and $|{\bf r}_{ij}|$ the
interparticle distance. We find that the non-monotonic character of
the curve is enhanced for $\tau>1$, while it tends to disappear for
$\tau <1$ and in particular it vanishes for $\tau \lesssim 0.9$. {In
the limit $\tau \rightarrow 0$ of purely repulsive interactions (not
shown) the system resembles the Coulomb case of Ref.~\cite{goldoni},
and accordingly we find that the MH lattice ceases to be the
groundstate configuration.}

The observations above confirm that the reentrant (non-monotonic) behavior of $T_m$ as a function of $\xi$ is due to the attractive character of the dipole-dipole interactions. % and thus it is reasonable to expect that it is not an artifact of the harmonic approximation.
This indicates that it is possible to alternate
solid and liquid phases by changing the interlayer distance or the
density of the molecules.
\\

%%%%%%%%%%%%%%%%%%%%%%%%%%%%%%%%%%%%%%%%%%%%%%%%%%%%%%%%%%%%

\section{conclusion}\label{sec:conc}

In this work we studied the structure, the stability and the melting
of a classical bilayer system of dipoles, polarized perpendicular to
the layers, as a function of the interlayer distance and the density
of dipoles in each {layer}. Using the Ewald summation technique, we
have computed the ground-state energy and the phononic spectrum of a
few physically-motivated lattice configurations, finding that the
ground state is always the matching hexagonal crystal configuration,
where two triangular lattices are stacked on top of each other, as
expected. Higher-energy configurations have been found to be
metastable in different regimes of interlayer distances and dipole
densities. These configurations may be realized using polar
molecules trapped in optical lattices of properly chosen {geometry}.

The main result of this work is that the anisotropic nature of the
dipole-dipole interaction potential profoundly affects the dynamical
properties of the bilayer system of dipoles, and its melting
behavior. In fact, on one hand we found that the attractive part of
the potential determines a non-monotonic dependence {on $\xi$} of
the longitudinal and transverse sound velocities  in the
ground-state configuration. On the other hand, we have shown that
the classical melting temperature of the bilayer crystal has an
interesting reentrant behavior as a function of $\xi$. This
reentrant behavior is due to the anisotropy of the dipole-dipole
interaction, and it entails that it is possible to alternate various
crystalline and liquid phases by changing $\xi$ at a fixed
temperature.

The present analysis is motivated by the recent developments in the
physics of cold molecular gases, which may provide for a physical
realization of these systems. In particular, the classical melting
of the bilayer crystalline phases may be realized in future in
setups where cold polar molecules are trapped in adjacent wells of a
one-dimensional optical lattice, and their dipoles are polarized by
a static electric field oriented parallel to the optical lattice.
Under appropriate trapping conditions \cite{Buechler07,Micheli07}, the
resulting {\it in-plane} dipole-dipole interactions are purely
repulsive, while inter-plane interactions can be repulsive or
attractive.%

\acknowledgments

The authors thank P. Zoller for stimulating discussions.
Lu acknowledges a supporting scholarship by Eurasia-Pacific Uninet
and the kind hospitality provided by the University of Innsbruck and
IQOQI of the Austrian Academy of Sciences.
This work is supported by the the National Natural Science
Foundation of China, the European Union project OLAQUI (FP6-013501-OLAQUI) and the Austrian Science Foundation (FWF).

\appendix

\section{rapidly convergent form of $\psi_0$ and $\psi_I$}\label{sec:appA}

The direct numerical computation of sums over lattice sites with long-range dipole-dipole interactions is in general impractical. Thus in the following we transform each sum into
rapidly convergent forms using Ewald method.\cite{goldoni,fisher} The detailed techniques are shown in the following. For the calculation of the rapidly convergent form of
the energy, at first we define the following two functions
\begin{subequations}
\begin{eqnarray}
\psi_0({\bf r},{\bf q}) &=&  e^{i {\bf q} \cdot {\bf r}} \sum_{j \ne
0}
   \frac{e^ {-i {\bf q} \cdot ({\bf R_j} + {\bf r})} }{|{\bf R_j} + {\bf r}|^3} ,  \\
\psi_{I}({\bf r},{\bf q}) &=&  e^{i {\bf q} \cdot {\bf r}} \sum_{j}
\Bigg( \frac{e^{i {\bf q} \cdot ({\bf R}_j + {\bf c} +
{\bf r})}}{|\tilde{\bf R}_j + {\bf r}|^3} \nonumber\\
&&+ \frac{- 3 ~l^2 ~e^{i {\bf q} \cdot
({\bf R}_j + {\bf c} + {\bf r})}}{|\tilde{\bf R}_j + {\bf r}|^5}  \Bigg) \nonumber \\
&=& \psi_{I1} - 3 ~l^2 \psi_{I2} ,
\end{eqnarray}
\end{subequations}
where
\begin{subequations}
\begin{eqnarray}
\psi_{I1} &=& \sum_{{j}} \frac{e^{i {\bf q} \cdot ({\bf R}_j + {\bf
c} +
{\bf r})}}{|\tilde{\bf R}_j + {\bf r}|^3} , \\
\psi_{I2} &=& \sum_{{j}} \frac{e^{i {\bf q} \cdot ({\bf R}_j + {\bf
c} + {\bf r})}}{ |\tilde{\bf R}_j + {\bf r} |^5} ,
\end{eqnarray}
\end{subequations}
with ${\bf R}_j\equiv{\bf R}_{\sigma,j}-{\bf R}_{\sigma,0}$ and
$|\tilde{\bf R}_j + {\bf r}|\equiv(|{\bf R}_j+{\bf c} + {\bf
r}|^2+l^2)^{1/2}$. Then $E_0$ and $E_I$ can be obtained from
\begin{subequations}
\begin{eqnarray}
E_0 &=& \lim_{{\bf r} \rightarrow 0} d^2 ~\psi_0({\bf r}, 0) , \\
E_I &=& \lim_{{\bf r} \rightarrow 0} d^2 ~\psi_{I}({\bf r}, 0) .
\end{eqnarray}
\end{subequations}
We use the identity based on the integral representation of the
gamma function
\begin{eqnarray}
\frac{1}{x^{2s}} = \frac{1}{\Gamma(s)} \int_0^{\infty} t^{s-1}
\exp(-x^2 t) \mathrm{d}t ,
\end{eqnarray}
with $s = 3/2, \Gamma(3/2)=\sqrt{\pi}/2$ for $\psi_0$ and
$\psi_{I1}$, $s = 5/2,\Gamma(5/2)=3\sqrt{\pi}/4$ for $\psi_{I2}$,
and the 2D Poisson summation formula
\begin{eqnarray}
&&\sum_{j} \exp\left(-|{\bf \rho} + {\bf R}_j |^2 t - i {\bf q}
\cdot
({\bf \rho} + {\bf R}_j)\right) \nonumber\\
&=& \frac{\pi}{L^2}~ t^{-1} \sum_{j} \exp\left( i {\bf G}_j \cdot
{\bf \rho} \right) \exp\left(- \frac{|{\bf G}_j + {\bf
q}|^2}{4t}\right) ,
\end{eqnarray}
where ${\bf G}_j = j_1 {\bf b}_1 + j_2 {\bf b}_2$ (with integers
$i,j$ ) is the two-dimensional vector in reciprocal lattice, $L^2 =
1/ n$ is the area per primitive cell.
Then $\psi_0({\bf r}, {\bf q})$ can be expressed by
\begin{eqnarray}
\psi_0({\bf r}, {\bf q}) &=& \frac{\pi}{L^2} \sum_{{j}} e^{i ({\bf
q} + {\bf G}_j) \cdot r} \frac{2}{\sqrt{\pi}} \int_0^{\alpha^2}
t^{-1/2} \nonumber\\
&&\exp\left( -\frac{|{\bf G}_j + {\bf q}|^2}{4 t} \right)
\mathrm{d}t - \frac{2}{\sqrt{\pi}} \int_0^{\alpha^2} t^{1/2}
e^{-|{\bf r}|^2 t} \mathrm{d}t \nonumber \\
&& + \sum_{{j} \ne 0} e^{-i {\bf q} \cdot {\bf R}_j}
\frac{2}{\sqrt{\pi}} \int_{\alpha^2}^{\infty} t^{1/2} e^{-|{\bf R}_j
+ {\bf r}|^2 t} \mathrm{d}t ,
\end{eqnarray}
where $\alpha$ is a small positive number. After using the
integration
\begin{eqnarray}
&&\int_{\alpha^2}^{\infty} t^{1/2} \exp(- |x|^2 t) \mathrm{d}t
\nonumber\\
&=& \frac{\sqrt{\pi}}{2 |x|^3} ~\mathrm{erfc}(\alpha |x|) +
\frac{\alpha \exp(- \alpha^2 |x|^2)}{|x|^2} \label{eq:int1} ,
\end{eqnarray}
and
\begin{eqnarray}
&&\int_0^{\alpha^2} t^{-1/2}~\exp \left(-\frac{|x|^2}{4t} \right)
\mathrm{d}t \nonumber\\
&=& \exp\left( -\frac{|x|^2}{4 \alpha^2} \right) \Bigg\{ 2 \alpha -
\exp\left(\frac{|x|^2}{4 \alpha^2}\right) \nonumber\\
&& \times |x| \sqrt{\pi} ~\mathrm{erfc}\left( \frac{|x|}{2\alpha }
\right) \Bigg\} ,
\end{eqnarray}
where the expression contains the complementary error function
$\mathrm{erfc}(z) = 1- \mathrm{erf}(z) = (2/\sqrt{\pi}) \int_0^{z}
e^{-t^2} \mathrm{d}t$, we obtain the final form of $\psi_0(r,q)$
%(Eq. \ref{psi0})
%
\begin{eqnarray}
\psi_{0}({\bf r}, {\bf q}) &=& \frac{\pi}{L^2} \sum_{j} e^{i ({\bf
q} + {\bf G}_j) \cdot {\bf r}} \Bigg\{ \frac{4 \alpha}{\sqrt{\pi}}
\exp \left( -\frac{|{\bf G}_j + {\bf q}|}{4
\alpha^2}\right) \nonumber\\
&&- 2 |{\bf G}_j + {\bf q}| ~{\rm erfc} \left( \frac{|{\bf G}_j +
{\bf q}|}{2 \alpha} \right) \Bigg\} \nonumber\\
&&+ \left[ \frac{2 \alpha e^{-\alpha^2 |r|^2}}{\sqrt{\pi} |{\bf
r}|^2} - \frac{\textrm{erf}(\alpha |{\bf r}|)}{|{\bf r}|^3}
\right] \nonumber \\
&& + \sum_{{j} \ne 0} e^{-i {\bf q} \cdot {\bf R}_j} \Bigg\{
\frac{{\rm erfc}(\alpha |{\bf R}_j + {\bf r}|)}{|{\bf R}_j +
{\bf r}|^3} \nonumber\\
&&+ \left( \frac{2 \alpha}{\sqrt{\pi}} \right) \frac{e^{-\alpha^2
|{\bf R}_j + {\bf r}|^2}}{|{\bf R}_j + {\bf r}|^2} \Bigg\}
\label{eq:psi0rap}.
\end{eqnarray}
Similarly,
\begin{eqnarray}
\psi_{I1} &=& \frac{\pi}{L^2} \sum_{j} e^{i ({\bf q} + {\bf G}_j)
\cdot r} e^{{\bf G}_j \cdot {\bf c}} \frac{2}{\sqrt{\pi}}
\int_0^{\alpha^2} t^{-1/2} \nonumber\\
&& \times \exp\left( \frac{|{\bf G}_j + {\bf q}|^2}{4 t} - l^2 t
\right) \mathrm{d}t  + \frac{2}{\sqrt{\pi}} \sum_{j} e^{-i {\bf q}
\cdot ({\bf R}_j +
{\bf c})} \nonumber\\
&& \times  \int_{\alpha^2}^{\infty} t^{1/2} \exp\left( -|\tilde{\bf
R}_j + {\bf r}|^2 t \right) \mathrm{d}t ,
\end{eqnarray}
by replacing $t$ by $w^{-2}$, the first integration of above
equation can be rewritten as
\begin{eqnarray}
\int_{1/\alpha}^{\infty} 2 w^{-2} \exp\left( -\frac{|{\bf G}_j +
{\bf q}|^2}{4} w^2 - \frac{l^2}{w^2} \right) \mathrm{d}w .
\end{eqnarray}
Using the integration
\begin{eqnarray}
&&\int_{1/\alpha}^{\infty} w^{-2} \exp\left( -\frac{|x|^2 w^2}{4}  -
\frac{y^2}{w^2} \right) \mathrm{d}w \nonumber\\
&=& \frac{\pi}{4 y} \Bigg[ e^{-|x|y} \mathrm{erfc}\left(
\frac{|x|}{2 \alpha} -\alpha y \right) \nonumber\\
&&- e^{|x|y} \mathrm{erfc}\left( \frac{|x|}{2 \alpha} +\alpha y
\right) \Bigg] ,
\end{eqnarray}
and Eq. \eqref{eq:int1} we have the form of $\psi_{I1}$
\begin{eqnarray}
\psi_{I1} &=& \frac{\pi}{L^2 l} \sum_{j} e^{i ({\bf q} + {\bf G}_j)
\cdot {\bf r}} e^{i {\bf G}_j \cdot {\bf c}} \Bigg[ e^{-|{\bf G}_j +
{\bf q}| l} \mathrm{erfc}\Bigg( \frac{|{\bf G}_j + {\bf q}|}{2
\alpha}
\nonumber\\
&& -\alpha l \Bigg) - e^{|{\bf G}_j + {\bf q}| l}
\mathrm{erfc}\left(
\frac{|{\bf G}_j + {\bf q}|}{2 \alpha} +\alpha l \right) \Bigg] \nonumber \\
&& + \sum_{j} e^{-i {\bf q} \cdot ({\bf R}_j + {\bf c})} \Bigg[
\frac{ \mathrm{erfc} \left( \alpha |\tilde{\bf R}_j + {\bf r}|
\right)}{|\tilde{\bf R}_j + {\bf r}|^3} \nonumber\\
&& + \left( \frac{2 \alpha}{\sqrt{\pi}} \right) \frac{\exp\left(
-\alpha^2 |\tilde{\bf R}_j + {\bf r}|^2 \right)}{|\tilde{\bf R}_j +
{\bf r}|^2} \Bigg].
\end{eqnarray}
In the same way, we transform $\psi_{I2}$ as
\begin{eqnarray}
\psi_{I2} &=& \frac{\pi}{3 L^2 l^3} \sum_{{j}} e^{i ({\bf q} + {\bf
G}_j) \cdot r} e^{i {\bf G}_j \cdot {\bf c}} \Bigg\{ -\frac{4
\alpha l}{\sqrt{\pi}} \nonumber\\
&& \times \exp\left(-\frac{|{\bf G}_j + {\bf q}|^2}{4 \alpha^2}
-\alpha^2 l^2 \right) + \bigg[
e^{-|{\bf G}_j + {\bf q}| l} \nonumber\\
&&\times \left( |{\bf G}_j + {\bf q}| l + 1 \right) \mathrm{erfc}
\left( \frac{|{\bf G}_j + {\bf q}|}{2 \alpha} - \alpha l \right)
\nonumber\\
&& + e^{|{\bf G}_j + {\bf q}| l} \left( |{\bf G}_j + {\bf q}| l - 1
\right) \mathrm{erfc}
\left( \frac{|{\bf G}_j + {\bf q}|}{2 \alpha} + \alpha l \right) \bigg] \Bigg\} \nonumber \\
&& + \sum_{{j}} e^{-i q \cdot ({\bf R}_j + {\bf c})} \Bigg\{ \frac{
\mathrm{erfc} \left(\alpha |\tilde{\bf R}_j + {\bf r}|
\right)}{|\tilde{\bf R}_j + {\bf r}|^5} \nonumber\\
&& + \left(\frac{2 \alpha}{3 \sqrt{\pi}} \right) \frac{3 + 2
\alpha^2 |\tilde{\bf R}_j + {\bf r}|^2}{|\tilde{\bf R}_j + {\bf
r}|^4} ~e^{ -\alpha^2 |\tilde{\bf R}_j + {\bf r}|^2 } \Bigg\} ,
\end{eqnarray}
where the integrations
\begin{eqnarray}
&& \int_{1/\alpha}^{\infty} w^{-4} \exp\left( -\frac{|x|^2 w^2}{4}
- \frac{y^2}{w^2} \right) \mathrm{d}w \nonumber \\
&=& \frac{1}{8 y^3} \Bigg\{ -4 \alpha ~y \exp\left( -\frac{|x|^2}{4
\alpha^2} -\alpha^2 y^2 \right) \nonumber\\
&& + \sqrt{\pi} \bigg[ e^{-|x|y} \left(|x|y + 1 \right)
\mathrm{erfc} \left( \frac{|x|}{2 y} - \alpha
y \right) \nonumber \\
&& + e^{|x|y} \left(|x|y - 1 \right) \mathrm{erfc} \left(
\frac{|x|}{2 y} + \alpha y \right) \bigg] \Bigg\} ,
\end{eqnarray}
and
\begin{eqnarray}
&&\int_{\alpha^2}^{\infty} t^{3/2} \exp\left(|x|^2 t \right)
\mathrm{d}t \nonumber\\
&=& \frac{3 \sqrt{\pi}}{4 |x|^5} ~\mathrm{erfc} (\alpha |x|) +
\frac{\alpha (3 + 2 \alpha^2 |x|^2)}{2 |x|^4} \nonumber\\
&&\times
\exp({-\alpha^2 |x|^2}),
\end{eqnarray}
are used. After simplifying, %we obtained $\psi_{I}$ (Eq.\ref{psii}).
we obtain the rapid convergent form of %$\psi_{0}({\bf r}, {\bf q})$
$\psi_{I}({\bf r},{\bf q})$ as
\begin{eqnarray}
\psi_{I}({\bf r},{\bf q}) &=& \frac{\pi}{L^2} \sum_{{j}} e^{i ({\bf
q} + {\bf G}_j) \cdot {\bf r}} e^{i {\bf G}_j \cdot {\bf c}} \Bigg\{
\frac{4 \alpha}{\sqrt{\pi}} \exp \Bigg( -\frac{|{\bf G}_j +
{\bf q}|^2}{4 \alpha^2} \nonumber\\
&&- \alpha^2 l^2 \Bigg) - |{\bf G}_j + {\bf q}| \bigg[ e^{-|{\bf
G}_j + {\bf q}| l} ~{\rm erfc}\Bigg(\frac{|{\bf G}_j + {\bf q}|}{2
\alpha}
\nonumber \\
&&-\alpha l \Bigg) + e^{|{\bf G}_j + {\bf q}| l} ~{\rm
erfc}\left(\frac{|{\bf G}_j +
{\bf q}|}{2 \alpha} +\alpha l \right) \bigg] \Bigg\} \nonumber\\
&&+ \sum_{{j}} e^{-i {\bf q} \cdot ({\bf R}_j + {\bf c})} \Bigg\{
\frac{{\rm erfc}\left( \alpha |\tilde{\bf R}_j + {\bf r}| \right)}{|\tilde{\bf R}_j + {\bf r}|^3} \nonumber \\
&& + \left( \frac{2 \alpha}{\sqrt{\pi}} \right) \frac{\exp \left(
-\alpha^2 |\tilde{\bf R}_j + {\bf r}|^2 \right)}{|\tilde{\bf R}_j +
{\bf r}|^2} \nonumber
\end{eqnarray}
\begin{eqnarray}
&&- 3 l^2 \bigg[ \frac{{\rm erfc}\left( \alpha |\tilde{\bf R}_j +
{\bf r}| \right)}{|\tilde{\bf R}_j + {\bf r}|^5} \nonumber\\
&& + \left( \frac{2 \alpha}{3 \sqrt{\pi}} \right) \frac{3 + 2
\alpha^2 |\tilde{\bf R}_j + {\bf r}|^2}{|(\tilde{\bf R}_j +
{\bf r}|^4} \nonumber\\
&& \times \exp \left( -\alpha^2 |\tilde{\bf R}_j + {\bf r} |^2
\right) \bigg] \Bigg\} \label{eq:psiirap}.
\end{eqnarray}
Therefore, the rapid convergent forms of $E_0$ and $E_I$ are
expressed as Eqs. \eqref{eq:e0rap} and \eqref{eq:eirap}.

%By using the relation $ 1/L^2 = n$.

%----------------------------------------------------------

%
\begin{figure}%[h]
\includegraphics[height=0.23\textheight]{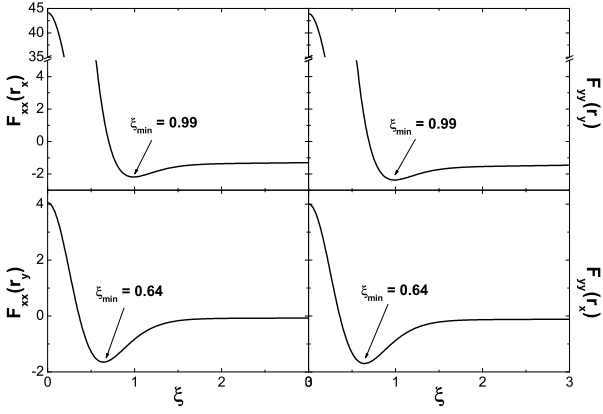}
\caption{(Color online) The {quantities} $F_{xx}^{r_x}$,
$F_{yy}^{r_y}$, $F_{xx}^{r_y}$ and $F_{yy}^{r_x}$ as functions of
$\xi$.} \label{fig:fig13:fxxyy}
\end{figure}

\section{interpretation of the minima of sound velocities for MH
configuration}\label{sec:appB}

For understanding the minimum of the LA and TA modes of the sound
velocities of MH configuration, we define the following functions
\begin{eqnarray}
F_{xx}^{r_x} &=& \lim_{u_x \rightarrow 0} \frac{\partial^2 E_I}{\partial u_x^2} ~r_x^2, \nonumber\\
F_{xx}^{r_y} &=& \lim_{u_x \rightarrow 0} \frac{\partial^2 E_I}{\partial u_x^2} ~r_y^2, \nonumber\\
F_{yy}^{r_x} &=& \lim_{u_y \rightarrow 0} \frac{\partial^2 E_I}{\partial u_y^2} ~r_x^2, \nonumber\\
F_{yy}^{r_y} &=& \lim_{u_y \rightarrow 0} \frac{\partial^2
E_I}{\partial u_y^2} ~r_y^2.
\end{eqnarray}
Where $E_I$ is the interaction potential between a {dipole} at
origin and the {dipoles} in the opposite layer, $u_{x(y)}$ is the
$x(y)$ component of the displacements of the origin {dipole} around
its equilibrium position, $r_{x(y)}$ is the $x(y)$ component of the
distance between the origin {dipole} and the {dipoles} in the
opposite layer.

The first two plots in Fig.~\ref{fig:fig13:fxxyy} describe
$F_{xx}^{r_x}$ and $F_{yy}^{r_y}$ as functions of $\xi$, which
indicate that the direction of vibration is along the direction of
propagation. We find that the minimum is located at the same value
of $\xi$ in the curves of $F_{xx}^{r_x}$ and $F_{yy}^{r_y}$ as that
in the plot of {$v_{\rm LA}$} [see Fig.~\ref{fig:fig05:svmh}(a)],
where {$\xi_{\rm min} \approx 1$}. The remaining two plots of Fig.
\ref{fig:fig13:fxxyy} {show} $F_{xx}^{r_y}$ and $F_{yy}^{r_x}$ vs
$\xi$, which express that the direction of vibration is
perpendicular to the direction of propagation, the value of $\xi$
for the minimum in this two curves is the same as that in {$v_{\rm
TA}$} [see Fig.~\ref{fig:fig05:svmh}(b)], where $\xi_{\rm min}=
0.64$. Due to the intrinsic property of interlayer dipole-dipole
interaction, the four functions above have minima at {$\xi_{\rm min}
\approx 1$} and $\xi_{\rm min}=0.64$ respectively.

%----------------------------------------------------------

\section{correlation functions for bilayer system}\label{sec:appC}

From the Fourier transformation we know that the $\gamma^{\rm th}$
component of the displacement vectors of a {dipole} at the origin
position {in layer $+$} and at the $h^{\rm th}$ nearest neighbor
position {in layer $\sigma$} are
\begin{eqnarray}
u_\gamma^{+} (0) &=& \frac{1}{\sqrt{N}} \sum_{{\bf q},j}
c_{\gamma}^{+} ({\bf q},j)
\rho_{\gamma}^{+}({\bf q},j) , \nonumber \\
u_\gamma^{\sigma} (h) &=& \frac{1}{\sqrt{N}} \sum_{{\bf q},j}
c_{\gamma}^{\sigma} ({\bf q},j) \rho_{\gamma}^{\sigma}({\bf q},j)
\exp(i {\bf q} \cdot {\bf R}_h^{\sigma}),
\end{eqnarray}
where {$\rho_{\gamma}^{\sigma}({\bf q},j)$ is the $\gamma$ component
of the eigenvector of $j^{\rm th}$ mode at ${\bf q}$ point in the
first Brillouin zone of the sublattice in layer $\sigma=\pm$}.
$c_{\gamma}^{\sigma}({\bf q},j)$ is the probability parameter of
$\rho_{\gamma}^{\sigma}({\bf q},j)$, {${\bf R}_h^{\sigma}$ is the
relative position of the $h^{\rm th}$ nearest neighbor dipole in
layer $\sigma$}. Making use of the relation $\langle
c_{\gamma}^{\sigma}({\bf q},j) c_{\gamma}^{\sigma'}({\bf q}',j')
\rangle = \left( k_B T/m \omega^2({\bf q},j) \right)
\delta_{q,q'}\delta_{j,j'}$, we can obtain the relative mean square
displacements between the two considered nearest neighbors
\begin{subequations}
\begin{align}
&\langle |u_{\gamma}^+(0) - u_{\gamma}^+(h)|^2 \rangle
\nonumber\\
=& \frac{2 k_B T}{N m} \sum_{{\bf q},j} \frac{1}{\omega^2({\bf
q},j)} \Bigg\{ \left[ \rho_{\gamma}^+({\bf q},j)^2 \right] \left( 1
- \cos {\bf q} \cdot
{\bf R}_h^+ \right) \Bigg\} , \label{eq:rmsd1} \\
%\end{eqnarray}
%
%\begin{eqnarray}
&\langle |u_{\gamma}^+(0) - u_{\gamma}^-(h)|^2 \rangle \nonumber \\
=& \frac{k_B T}{N m} \sum_{{\bf q},j} \frac{1}{\omega^2({\bf q},j)}
\Bigg\{ \left[ \rho_{\gamma}^+({\bf q},j)^2 + \rho_{\gamma}^-({\bf
q},j)^2 \right]
\nonumber\\
& - 2 \left[ \rho_{\gamma}^+({\bf q},j) \rho_{\gamma}^-({\bf q},j)
\right] \cos\left( {\bf q} \cdot {\bf R}_h^- \right) \Bigg\} ,
\label{eq:rmsd2}
\end{align}
\end{subequations}
where $u_{\gamma}^{+}(0)$ and  $u_{\gamma}^{+(-)}(h)$ are the
$\gamma$ component of the original and the {$h^{\rm th}$} nearest
neighbor {dipoles} in the $+(-)$ layer, $m$ is the mass of {the
dipoles}, $k_B$ is the Boltzmann constant, $\omega({\bf q},j)$ is
the phonon frequency of {$j^{\rm th}$} mode at ${\bf q}$ point in
the first Brillouin zone. After the summation over the {$\gamma^{\rm
th}$} components and over the nearest neighbor sites, we can finally
obtain the expressions Eqs. \eqref{eq:cf1} and \eqref{eq:cf2}.

%%%%%%%%%%%%%%%%%%%%%%%%%%%%%%%%%%%%%%%%%%%%%%%%%%%%%%%%%%%%%%%%

%\begin{references}

\end{document}